\documentclass[11pt,a4paper]{article}
\usepackage{jheppub}

\usepackage[utf8]{inputenc}
\usepackage{lmodern}

\usepackage{lineno}

\usepackage{float} 
\usepackage{subfig} 
\usepackage{amsmath} %

\usepackage{graphicx}%
\usepackage{tabularx,ragged2e}%

\newcolumntype{Y}{>{\centering\arraybackslash}X}

\newcommand{\eqs}[1]{\begin{equation} \begin{split} #1\end{split} \end{equation} }

\def\ie{{\it i.e.}}
\def\eg{{\it e.g.}}

\def\GeV{{\rm GeV}}
\def\GeV2{{\rm GeV}^2}

\providecommand{\pp}{$pp$}
\providecommand{\pp}{$pd$}
\providecommand{\pA}{$pA$}

\providecommand{\pPb}{$p$Pb}
\providecommand{\pPb}{$p\,$Pb}
\providecommand{\Pbp}{Pb$p$}
\providecommand{\Arp}{Ar$p$}
\providecommand{\Op}{O$p$}

\providecommand{\AaBb}{$AB$}

\providecommand{\AuAu}{AuAu}

\providecommand{\InIn}{InIn}
\providecommand{\PbPb}{PbPb}
\providecommand{\gaga}{{\gamma\,\gamma}}

\newcommand{\sqrtsgh}{\sqrt{s_{{\gamma h}}}}
\newcommand{\sqrtsnn}{\sqrt{s_{NN}}}
\newcommand{\snn}{s_{_{\ensuremath{\it{NN}}}}}

\newcommand{\ce}[1]{Eq.~(\ref{#1})}
\newcommand{\cf}[1]{{Fig.~\ref{#1}}}
\newcommand{\ct}[1]{{Table~\ref{#1}}}

\newcommand{\Lumi}{\mathcal{L}}

\newcommand{\re}{\mathrm{Re}\,}

\newcommand{\gev}{\,{\rm GeV}}

\title{Lepton-pair production in ultraperipheral collisions at AFTER@LHC}
\author[a]{J.P. Lansberg}
\author[b]{L. Szymanowski}
\author[b]{J. Wagner}
\affiliation[a]{IPNO, Universit\'e Paris-Sud, CNRS/IN2P3, F-91406, Orsay, France}
\affiliation[b]{National Centre for Nuclear Research (NCBJ), Ho\.{z}a 69, 00-681, Warsaw, Poland}

\emailAdd{Jean-Philippe.Lansberg@in2p3.fr}
\emailAdd{Lech.Szymanowski@fuw.edu.pl}
\emailAdd{Jakub.Wagner@ncbj.gov.pl}

\abstract{
We investigate the potentialities offered by the study of lepton-pair production in ultraperipheral collisions at a 
fixed-target experiment using the proton and ion LHC beams. In these collisions, exclusive or semi-exclusive 
lepton-pair production can be used as luminosity monitor as well as a check of the equivalent-photon approximation, 
via the measurement of the Bethe-Heitler cross section. It can also serve  as a probe of the inner hadron structure
via the measurement of the lepton-pair azimuthal asymmetry which is sensitive to the timelike virtual Compton scattering.
We also briefly discuss the possibility offered by the study of $\eta_c$ production. 
Finally, we outline the possibilities for lepton-pair production by Pomeron-Odderon fusion in exclusive $pp$ and $pA$ collisions.
}
\date{\today}

\begin{document}

\maketitle

\section{Introduction.}
With the advent of RHIC and the LHC, the first experimental studies of Ultra-Peripheral Collisions (UPC) have successfully
been carried out. The STAR collaboration first measured the $\rho_0$ production cross section in AuAu UPC~\cite{Adler:2002sc}, then
they measured the di-electron production cross section in UPC~\cite{Adams:2004rz}. A little later the PHENIX collaboration
released their first study of the $J/\psi$ cross section as well as high-mass di-electron production~\cite{Afanasiev:2009hy}. More recently, 
further studies have been carried out at the LHC by the ALICE collaboration~\cite{Abelev:2012ba,Abbas:2013oua}.

Attempts to isolate UPC in InIn collisions in the fixed-target mode at the SPS have been made\footnote{P. Ramalhete,  \href{https://twiki.cern.ch/twiki/pub/Main/CarlosLourenco/Thesis.pdf}{PhD. thesis}, April 2009.}. 
They were not conclusive, most probably  because of the limited nucleon-nucleon centre-of-mass (cms) energy, on the order 
of 20 GeV, resulting in typical photon-nucleon cms energies below 3 GeV.

In this context, we investigate the possibility to study lepton-pair production in ultraperipheral 
collisions at a fixed-target experiment using the proton and ion LHC beams~\cite{Brodsky:2012vg} -- referred 
thereafter to as AFTER@LHC. In such collisions, one can investigate specific reactions where one of the colliding
particle serves as a (coherent) emitter of a photon and the other serves as a target. Such photon-hadron  collisions
can be (semi-)exclusive, resulting, for instance, in the sole production of a lepton pair. This process can be used to monitor 
the experiment luminosity since it mainly comes from the Bethe-Heitler (BH) process~\cite{Bethe:1934za}, whose cross section 
is well-known. By looking at the target rapidity region, it can also be used to scan the domain of validity of the 
Equivalent-Photon Approximation (EPA). It can also be used to probe the inner structure of the target, through the interference 
between the BH process and the Timelike Compton Scattering (TCS)~\cite{Berger:2001xd,Pire:2011st,Moutarde:2013qs}. 
Such an interference -- measurable via the analysis of the azimuthal anisotropy -- indeed involves 
contributions from the Generalised Parton Distributions (GPD)~\cite{Mueller:1998fv,Ji:1996ek,Radyushkin:1997ki,Collins:1998be,Diehl:2003ny,Belitsky:2005qn,Boffi:2007yc,Guidal:2013rya}.

The structure of this article is as follows. In section 2, we present the main characteristics of the UPCs and the corresponding
photon fluxes in a fixed-target mode on the LHC beams. In section 3, we briefly discuss the cross sections for 
production of lepton pair via the BH process. In section 4, we discuss how the contribution from TCS 
can be extracted and how they can help to unravel information about the inner proton structure.
In section 5, we briefly discuss the potential competing hadronic process resulting from photon-odderon fusion.
Finally, we present our outlooks and conclusions.

\section{Ultraperipheral collisions in a high-energy fixed-target experiment}

\subsection{Generalities on photon-induced reactions in ultraperipheral collisions}

Relativistically moving charged hadrons are accompanied by electromagnetic fields 
which can effectively be used as quasi-real-photon beams. At very high
energies, these photons are energetic enough to initiate hard reactions,
just as in lepton-proton colliders.

The virtuality, $q^2=-Q^2$, of these photons is small, $Q^{2} \lesssim 1/R^{2}$, where $R$ 
is the radius of the charge. More precisely,  $Q \lesssim$~0.28~GeV for protons ($R\approx$~0.7~fm) 
and $Q \lesssim 0.06$~GeV for nuclei ($R_A\approx 1.2\,A^{1/3}$~fm) with a mass number $A>$~16. These
photons, which are emitted coherently, are  almost on mass shell, and their emission 
can be theoretically treated in the EPA (see \eg~\cite{Budnev:1974de}).

Seen from a target at rest, the energy of these photons can become significant if the
energy of the moving charge, \ie\ the beam energy, becomes ultra-relativistic, as at the LHC. At rest, the coherent photon cloud
of an heavy ion is on the order of 30 MeV. Boosted a few thousand times ($\gamma_{\rm Pb}\simeq 2940$), these photons
have an energy close to 100 GeV in the laboratory frame. It is of course much less than what can be achieved
at the LHC in the collider mode, but close to the experimental condition at RHIC with colliding beams of 100 GeV.
It is anyhow enough to produce hard dileptons as well as vector mesons.

The energy spectrum of these photons depends on the boost with respect to the observer 
as well as the impact parameter $b$ -- it is understood that the observer or the probe is outside the charge distribution.
Using the EPA method, one gets~\cite{Baltz:2007kq} that the flux as function of the photon momentum $k$, of $b$ and $\gamma$ (the Lorentz factor of the hadron -- or nucleus -- in the frame where $k$ is measured)
reads~
\begin{eqnarray}
\frac{dn}{dkd^2b} = \frac{Z^2\alpha_{\rm em} \omega(b,k)^2}{\pi^2 k b^2}
\left[K_1^2(\omega(b,k))+\frac{1}{\gamma^2}K_0^2(\omega(b,k)) 
\right],\label{eq:dndkd2b}
\end{eqnarray}
where $\alpha_{\rm em}$ is the QED coupling, $Z$ is the nucleus charge, $\omega(b,k)=kb /\gamma$ and
$K_{1,2}$ are modified Bessel functions of the second kind.

\begin{table}[htb!]
\begin{center}
\caption[]{Relevant parameters for  $AB$ UPCs at  AFTER@LHC, at RHIC and at SPS:
(i) nucleon-nucleon cms, $\sqrtsnn$ (ii)  luminosity, $\Lumi_{AB}$, 
(iii-iv) colliding hadron energies, $E^{\rm lab}_{A,B}$, in the laboratory frame, 
(v) Lorentz factor between the colliding hadron rest frame and cms, $\gamma=\sqrtsnn/(2\,m_N)$,  
(vi) Lorentz factor between both colliding hadron rest frames, $\gamma=\snn/(2\,m^2_N)$,
(vii-viii) inverse of the colliding hadron effective radii (giving the typical photon cloud energy in the emitter rest frame) 
(ix-x) photon ``cutoff energy'' in the target (resp. projectile) rest frame, $E_{\gamma~\rm max}^{\rm B\ rest}$ (resp. $E_{\gamma~\rm max}^{\rm A\ rest}$)
(xi-xii) ``maximum'' photon-nucleon cms energy where $A$ (resp. $B$) is the photon emitter, $\sqrt{s_{\gamma_N}^{\rm max}}$ (resp. $\sqrt{s_{_N\gamma}^{\rm max}}$)
(xiii) photon ``cutoff energy'' in the cms, $E_{\gamma~\rm max}^{\rm cms}$, with both $A$ and $B$ emitting a photon coherently
(xiv) ``maximum'' photon-photon cms, $\sqrt{s_{\gaga}^{\rm max}}$.
}
\label{tab:UPC-parameters}
\scriptsize
\begin{tabularx}{\textwidth}{YYYcYYYYYYYYYY}
\hline\hline 
System  & target thickness & $\sqrt{s_{{NN}}}$ & ${\cal L}_{AB}$\protect\footnotemark 
   & $E^{\rm lab}_{A}$ &  $E^{\rm lab}_{B}$ & $\gamma^{\rm cms}$ &  $\gamma^{\rm A \leftrightarrow B}$ & $\frac{\hbar c}{R_A+R_B}$ & $E_{\gamma~\rm max}^{\rm A/B\ rest}$  & $\sqrt{s_{\gamma_N}^{\rm max}}$  & $E_{\gamma~\rm max}^{\rm cms}$ & $\sqrt{s_{\gaga}^{\rm max}}$  \\
 &(cm)      & (GeV) & (pb$^{-1}$yr$^{-1}$) & (GeV) & (GeV) & $\big(\frac{\sqrt{\snn}}{2m_N}\big)$ & $\big(\frac{\snn}{2m^2_N}\big)$ & (MeV) & (GeV) & (GeV)  & (GeV)& (GeV)   \\ \hline
AFTER@LHC\\
\pp   & 100  & 115 & $2.0 \times 10^4$   & 7000    & $m_N$     & 61.0 & 7450 & 141    & 1050  & 44   & 8.6  & 17  \\ 
\pPb  & 1    & 115 & 160   & 7000    & $m_N$     & 61.0 & 7450 & 25.3   & 188   & 19   & 1.5  & 3.1 \\ 
$pd$  & 100  & 115 & $2.4 \times 10^4$   & 7000    & $m_N$     & 61.0 & 7450 & 69.5   & 517   & 31   & 4.2  & 8.5 \\  
\PbPb & 1    & 72  & $7. \times 10^{-3}$  & 2760    & $m_N$     & 38.3 & 2940 & 13.9   & 40.7  & 8.8  & 0.53 & 1.1 \\  
\Pbp  &100   & 72  & 1.1                 & 2760    & $m_N$     & 38.3 & 2940 & 25.3   & 74.2  & 12   & 0.97 & 1.9 \\ 
\Arp  &100   & 77  & 1.1                 & 3150    & $m_N$     & 40.9 & 3350 & 41.1   & 138   & 16   & 1.7  & 3.4 \\ 
\Op   & 100  & 81  & 1.1                 & 3500    & $m_N$     & 43.1 & 3720 & 53.0   & 197   & 19   & 2.3  & 4.6 \\ \hline 
RHIC\\
\pp   & n/ap & 200 & 12                   & 100    & 100       & 106   & 22600 & 141   & 3190  & 77   & 15   & 30  \\ 
\AuAu & n/ap & 200 & $2.8 \times 10^{-3}$  & 100    & 100       & 106   & 22600 & 14.2  & 320   & 25   & 1.5  & 3.0\\ \hline
SPS\\
\InIn & n/av & 17   & n/av                    & 160    & $m_N$      &  9.23 & 170  & 16.9   & 2.87  & 2.4  & 0.16 & 0.31\\ 
\PbPb & n/av & 17   & n/av                    & 160    & $m_N$      &  9.23 & 170  & 13.9   & 2.36  & 2.1  & 0.13 & 0.26\\ \hline\hline
\end{tabularx}
\end{center}
\end{table}

\footnotetext{For \Arp\ and \Op\ luminosity with AFTER@LHC, we conservatively assumed the same extracted flux of Ar and O as for Pb, \ie\ $2 \times 10^5$ Pb/s.
See also~\cite{Brodsky:2012vg,Lansberg:2012wj,Rakotozafindrabe:2013au,Massacrier:2015nsm}}

Unless $b$ is smaller\footnote{Otherwise, (i) one cannot consider the entire nucleus charge $Z$, (ii) hadronic interactions may be more important
that the photon-induced ones, (iii) the probably for the colliding objects to break-up may also be important.} than $R$, 
the strong suppression of the flux by the Bessel functions when $\omega$ gets of the order of unity implies that
$k$ should be smaller than inverse radius times the Lorentz factor $\gamma$. The larger $\gamma$ is and the smaller the emitter is, the harder the flux is. 
We also note that the energy spectrum $k dn/dk$ is constant for $k/\gamma$ fixed.
To fix the idea, one usually considers a maximum photon momentum $k^{\rm max} \simeq \frac{\hbar c}{R_{\rm emitter}}$ below which
the emission are likely coherent and therefore characterised by a flux proportional to $Z^2$. This quantity should  not thus be considered
as a sharp cut-off above which photon emissions are forbidden. 
If the photon is considered in the hadron cms, one has $\gamma = \sqrt{s_{NN}}/(2m_N)(\equiv\gamma^{\rm cms})$.
If the photon is considered in the target $B$ rest frame,  $\gamma= s_{NN}/(2m_N^2)(\equiv\gamma^{A\to B})$. 
\ct{tab:UPC-parameters} summarises the relevant parameters characterising ultra-peripheral collisions at AFTER@LHC, at RHIC and a SPS in fixed-target mode.

For comparisons with photon-induced reactions in the more conventional lepton-hadron collisions, 
it is usually more instructive to look at the maximum of the $\gamma N$ cms energy, 
$\sqrt{s_{\gamma N}^{\rm max}}$. To do so, we ``boost''\footnote{In fact, the procedure looks more as if the emitter is Lorentz contracted as $1/\gamma$, rather than the
photon momentum boosted. The results are however similar once one considers the emitted photon as slightly off-shell, 
with a momentum, in the rest frame of the emitter, as $(k^{\rm max},0,0,0)$.} $k^{\rm max}$ in the ``target'' nucleon rest frame, 
to obtain $E_{\gamma~\rm max}^{\rm N\ rest}=\gamma^{A\to B} k^{\rm max} $,
 where $\sqrt{s_{\gamma p}}=\sqrt{2  E^{\rm max}_{\rm N\ rest} m_N}=\sqrt{s_{NN}k^{\rm max}/m_N}$. 
We note that the photon-energy ``cut-off'' obtained with these dimensional arguments for Pb$p$, \ie\ 74 GeV, 
is remarkably close to the peak in the energy spectrum obtained using a more realistic model in a recent study of 
the Bremsstrahlung spectrum of ions in AFTER@LHC~\cite{Mikkelsen:2015dva}, \ie\ 80 GeV.

\subsection{Photon fluxes}

Taking into account the smallest possible impact parameter for a given colliding system, \pp, \pA\ or \AaBb, 
as well as the charge distribution through a form factor in the proton case, one obtains different formulae 
for the flux integrated in $b$. In fact, the photon fluxes do not formally factorise since
$b_{min}$ depends on the radius of both colliding objects, except in $ep$ collisions, where one can reasonably
neglect the electron radius.

Along these lines, one should normally have for \pp\ collisions, $b_{\rm min}\simeq 2\times R_p$; 
for \pA\ collisions, $b_{\rm min}\simeq R_p+R_A$; and for $AB$ collisions  $b_{\rm min}\simeq R_A+R_B$.
Whereas it is acceptable to approximate $R_p+R_A$ to $R_A$, it does not seem justifiable
to use $R_{\rm Pb}$ for PbPb collisions, for instance. In addition, in \pA\ collisions, it is also 
problematic to use a different $b_{\rm min}$ when one considers the proton emission
or the ion emission. In both cases, one should use $R_p+R_A$, or perhaps $R_A$.

Integrating \ce{eq:dndkd2b} over $b$, one has~\cite{Baltz:2007kq}
\begin{eqnarray}
\frac{dn}{dk} = \frac{2 Z^2\alpha_{\rm em}}{\pi k}
\Bigg[
\omega(b_{\rm min},k)K_0\big(\omega(b_{\rm min},k)\big)K_1\big(\omega(b_{\rm min},k)\big) \nonumber\\
-\frac{{\omega(b_{\rm min},k)}^2}{2}
\big(K_1^2\big(\omega(b_{\rm min},k)\big)-K_0^2\big(\omega(b_{\rm min},k)\big)\Big)
\Bigg].
\end{eqnarray}

To avoid any confusion with the choice of the frame, it is useful to work with the momentum fraction or 
light cone coordinate, $x_\gamma=k/p_h\simeq k/(\gamma M_N)$, where $p_h$ is the momentum of the hadron emitting the photon. 
One trivially obtains
\eqs{
\frac{dn}{dx_\gamma} = \frac{k}{x_\gamma}\frac{dn}{dk}\Big|_{\omega_{pA} = x_\gamma M_p b_{min}}.
}  
The relation between the (differential) hadron-hadron cross section, $(d)\sigma^{h_Ah_B}$,
and the (differential) cross section for a photo-hadron scattering ($h_A$ or $h_B$), $(d)\sigma^{\gamma h_{A,B}}$, is naturally given by the 
following convolution with the photon flux 
\eqs{\label{eq:intWW}
d\sigma^{h_Ah_B}=&\int dk_\gamma \Big( \frac{dn^{h_A}}{dk_\gamma} \, d\sigma^{\gamma h_B}(s_{\gamma h_B}(k_\gamma))+
 \frac{dn^{h_B}}{dk_\gamma} \, d\sigma^{\gamma h_A}(s_{\gamma h_A}(k_\gamma))\Big)\\
= &\int dx_\gamma \Big( \frac{dn^{h_A}}{dx_\gamma} \, d\sigma^{\gamma h_B}(s_{\gamma h_B}(k_\gamma(x_\gamma))+\frac{dn^{h_B}}{dx_\gamma} \, d\sigma^{\gamma h_A}(s_{\gamma h_A}(k_\gamma(x_\gamma)))\Big).}
By analogy with the parton model formulae, one can thus write:
\eqs{
\varphi_\gamma(x_\gamma)=\frac{dn}{dx_\gamma},}
and interpret the latter as an equivalent-photon PDFs off the hadron (or the ions) by
leaving the emitter intact.

\begin{figure}[hbt!]
\begin{center}
  \includegraphics[width=0.75\columnwidth]{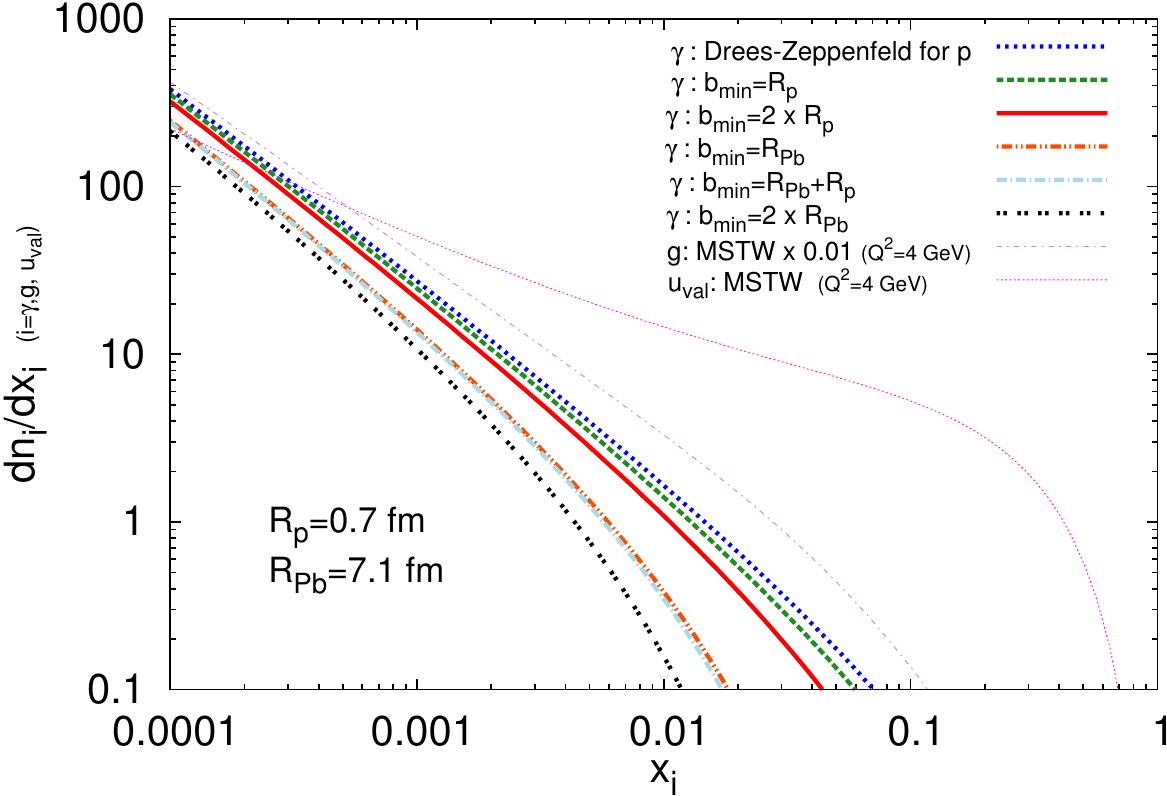} 
\caption{$\frac{dn}{dx}$ for photons from protons (blue dot for Drees-Zeppenfeld, green dash for $b_{\rm min}=R_p$ and red solid for $b_{\rm min}=2R_p$)
 and from Pb (orange double dot dash for $b_{\rm min}=R_{\rm Pb}$, light blue dot dash for $b_{\rm min}=R_{\rm Pb}+R_p$ and black double dot for $b_{\rm min}=2R_{\rm Pb}$ )  divided by $Z_{\rm Pb}^2$. 
These are compared to the gluon (thin pink long-dash, divided by 100) and $u$ quark (thin purple dot-dash) MSTW PDF in the proton.}
\label{fig:flux_x}
\end{center}
\end{figure}

In the case where the emitter is a proton, one could think that it is more accurate 
to take into account the spatial distribution of the charges through a form factor, which leads~\cite{Drees:1988pp} to
\begin{eqnarray}
\frac{dn}{dx_\gamma} = \frac{\alpha}{2\pi x_\gamma}\left(1 + (1-x_\gamma)^2\right)
\left[
\ln c-\frac{11}{6} +\frac{3}{c}-\frac{3}{2c^2}+\frac{1}{3c^2}
\right],
\end{eqnarray}
where $c = 1+\frac{0.71 \gev^2}{Q_{min}^2}$, and $Q_0^2 = M_p^2 x_\gamma^2$. This is only strictly relevant for 
$ep$ collisions and at rather large $x_\gamma$. In $pp$ or $pA$ collisions, one indeed needs to take into account
the radius of the other colliding object. The impact parameter $b$ is therefore not getting close to $R_p$.
We will refer to this choice by ``Drees-Zeppenfeld''.

\cf{fig:flux_x}
shows a comparison (up to the charge factor $Z^2$) between the equivalent flux
from a proton and from a lead ion. For $x_\gamma < 10^{-3}$, the spectra show a similar
behaviour, with a magnitude differing by less than a factor 2. At larger  $x_\gamma$, the proton
spectrum is clearly harder because of the smaller proton size. This is the expected behaviour.

One could also consider the case where the proton emitting the photon breaks apart.
In this situation, the photon is in fact radiated by the quarks and the corresponding photon PDF 
can then approximated by~\cite{Drees:1994zx}
\eqs{
\varphi_\gamma^{\rm break-up}(x,Q^2)=&\\
\frac{\alpha}{2\pi} \log\frac{Q^2}{Q^2_0} &\sum_q \int_x^1 \frac{dy}{y} P_{\gamma q}(x/y) \;\left[q(y,Q^2)+\bar q(y,Q^2)\right],
\label{eq:flux_p_inel}
}
with $P_{\gamma q}(z)=e^2_q \,(1+(1-z)^2)/z$, $Q^2_0$ is an energy cut-off, and $q(x,Q^2)$ is the quark PDFs in the proton.
We will not consider this possibility further in this study, although such a process could contribute to the semi-exclusive case
with still a large rapidity gap between the dilepton and the proton emitter. Another possibility, which we 
will discuss in section~\ref{sec:odderon} is to have an elastic {\it hadronic} reaction via the exchange of a pomeron 
(or possibly an odderon). 

\subsection{Fluxes and the rapidity dependence of the produced particles}

\subsubsection{Single-photon case}

If we consider the {\it total} cross section, $\sigma^{\gamma h}$, to photo-produce a particle
of mass $Q$, we note that it can only be function of $s_{\gamma h}$ and $Q^2$ since it is 
already integrated on the final state variables. However, nothing prevents us to
perform a change of variable in~\ce{eq:intWW} from $k_\gamma$ to a final state variable
of the $\gamma$-hadron process, for instance the rapidity $y$ of the produced particle, keeping the
other fixed\footnote{In the following, we denote the 4-momentum of this particle $q=(q_0,\vec q_T, q_z)$ and $q_T=|\vec q_T|$.}. 
By momentum conservation, $y$ would enter\footnote{
Defining $x_\gamma=s_{\gamma h}/s$, one indeed gets $x_\gamma=x_\gamma(y,q^2_T,Q,\epsilon)$ where $\epsilon=\pm 1$ when the projectile (target) is the photon emitter.
See the appendix \ref{appendix-1} for details.}
 via $k_\gamma(y)$ in $s_{\gamma h}$

Indeed, in our case, it is clearly instructive, in order  to understand where the produced particle 
by photon-induced collisions fly, to consider the flux as a function of the 
particle rapidity $y$ in the laboratory frame. For instance, we anticipate~\cite{Brodsky:2012vg} that
rapidities (in the laboratory frame) from -4 to +1 should be accessible.

\begin{figure}[h]
\begin{center}
\subfloat[\, proton on Pb]{\includegraphics[height=0.4\textwidth]{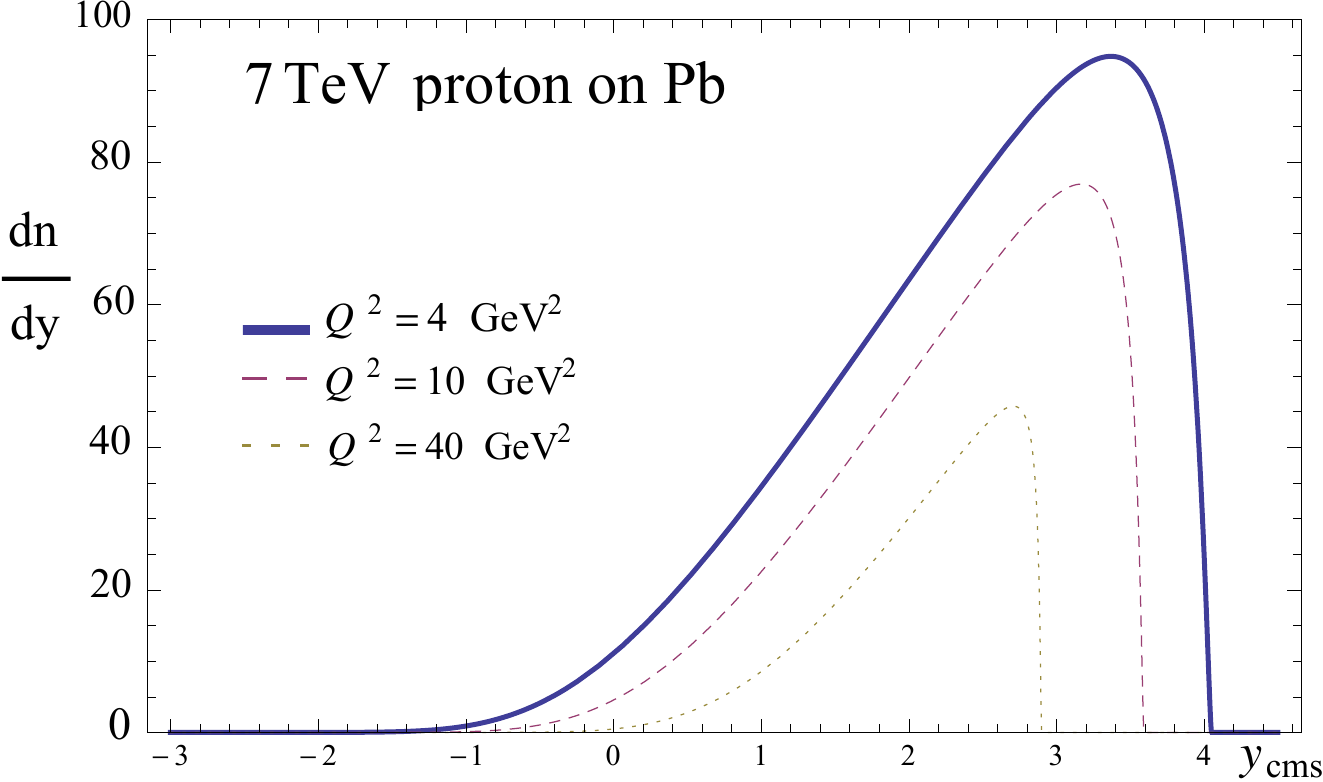}\label{fig:flux_y_p_on_Pb}}
\vspace*{-0.2cm}\\
\subfloat[\, Pb on H]{ \includegraphics[height=0.4\textwidth]{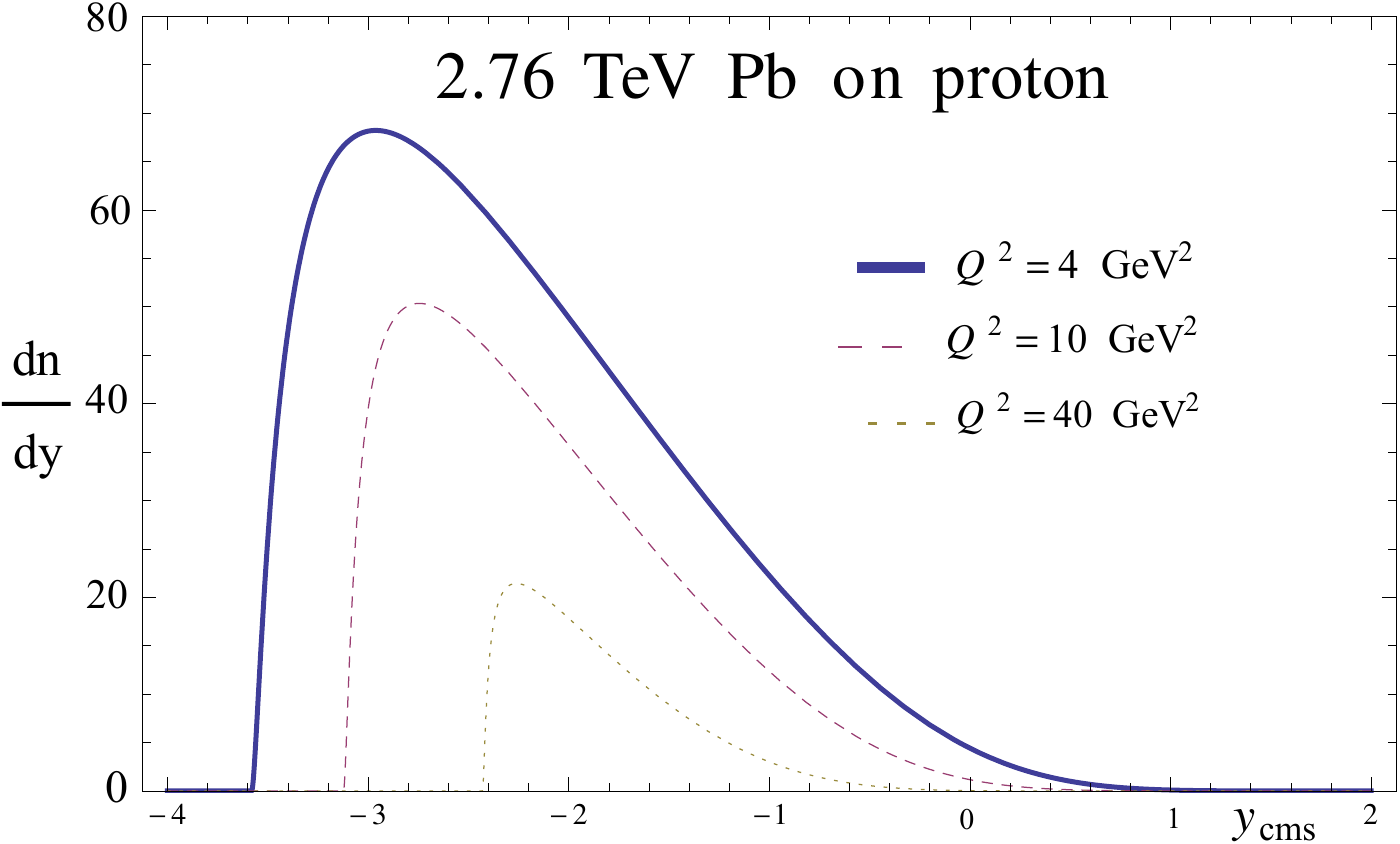}\label{fig:flux_y_Pb_on_p}}
\vspace*{-0.2cm}\\
\subfloat[\, proton on H]{\includegraphics[height=0.4\textwidth]{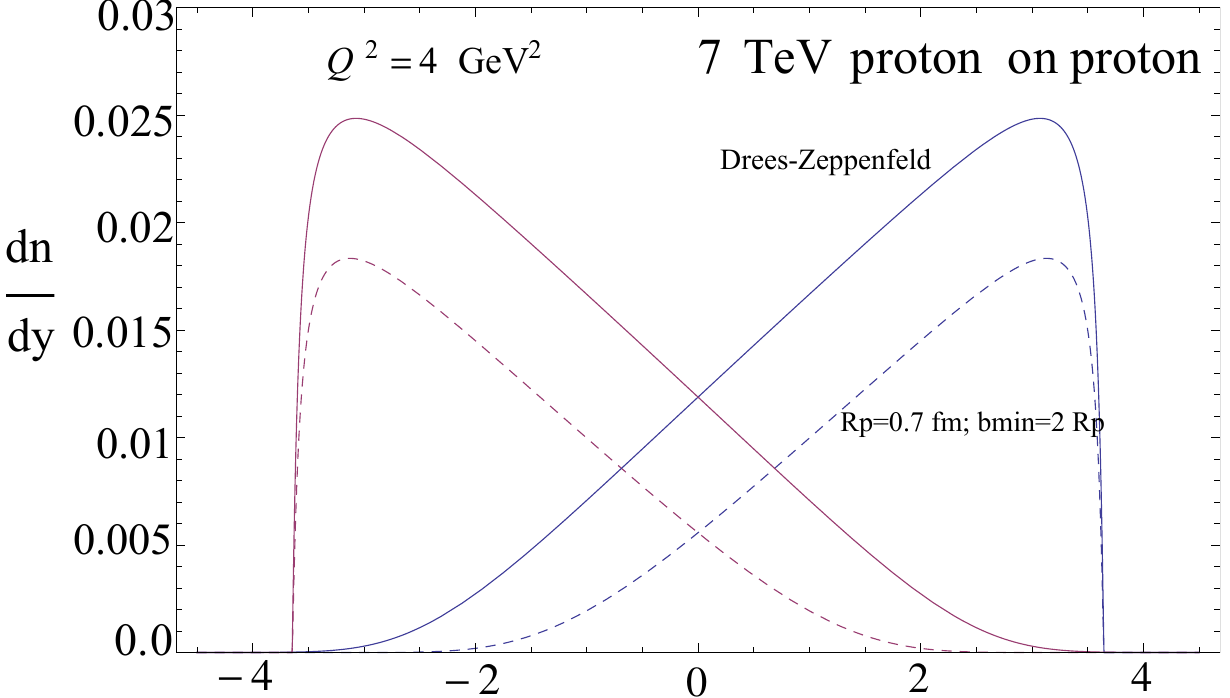}\label{fig:flux_y_p_on_p}} 
\vspace*{-0.2cm}\caption{$\frac{dn}{dy}$ for the proton run on Pb nucleon target (a), and for Pb run on H target (b), and (c) for p-p run. 
The flux is calculated for the specified 
$Q^2$ at  $q_T=0$ (\eg~$t=t_0$).\vspace*{-0.5cm}}
\label{fig:flux_y}
\end{center}
\end{figure}

For fixed $q_T$ and $Q^2$, $k_\gamma(y)$ is also fixed. From \ce{eq:intWW}, one trivially obtains 
\eqs{
d\sigma^{hh}&=\int dk_\gamma \ \Bigg( \frac{dn^{h_A}}{dk\gamma} \, d\sigma^{\gamma h_B}(s_{\gamma h_B}(k_\gamma))+
\frac{dn^{h_B}}{dk\gamma} \, d\sigma^{\gamma h_A}(s_{\gamma h_A}(k_\gamma)) \Bigg)\\
&= \int dy \ \Bigg(  
\frac{dn^{h_A}}{dy} \, d\sigma^{\gamma h_B}(s^A_{\gamma h_B}(y))
+
\frac{dn^{h_B}}{dy} \, d\sigma^{\gamma h_A}(s^B_{\gamma h_A}(y))
\Bigg),}
where $s^{(A,B)}_{\gamma h_{(B,A)}}(y)\equiv s_{\gamma h_{(B,A)}}\big(k_{\gamma \text{ from (A,B)}}(y)\big)$
and therefore
\eqs{\label{eq:rap-dep-dndy}
\frac{d\sigma^{hh}}{dy}= \frac{dn^{h_A}}{dy} \, d\sigma^{\gamma h_B}(s^A_{\gamma h_B}(y))
+\frac{dn^{h_B}}{dy} \, d\sigma^{\gamma h_A}(s^B_{\gamma h_A}(y))
.}

We therefore find it instructive to plot $\frac{dn}{dy}$ (for fixed $q_T$ and $Q^2$) for different 
configurations: \cf{fig:flux_y_p_on_Pb} shows the case of 7 TeV protons on lead, the lead being the photon 
emitter; \cf{fig:flux_y_Pb_on_p} shows the case of 2.76 TeV lead on protons, the lead being the photon 
emitter\footnote{In this case, as for (a), the photon emission by a lead ion can, in principle, be 
tagged with a neutron emission. In addition, the probability for the proton to be the emitter is, in practice, 
negligible ($Z^2$ suppressed at the same 
$|y_{cms}|$). We therefore do not show the curve for this possibility.}; \cf{fig:flux_y_p_on_p} shows the case of 
7 TeV protons on proton (i.e. hydrogen), where both can be the photon emitter (note that the corresponding fluxes 
cannot simply be summed).

In the first case (\cf{fig:flux_y_p_on_Pb}), the flux is maximum in the forward region, which corresponds to a 
soft (coherent) emission by the lead target. In general, the particle tends to be emitted in the ``hadron-receiver'' 
region since the photon momentum is very small (here $y_{\rm receiver = beam}=4.8$). To be more precise, the boost -- or 
rapidity difference $\Delta y$ -- between the photon-hadron and hadron-hadron cms,  
$\Delta y^{\gamma h_{A,B}}_{h_Ah_B} = y_{\rm cms}^{h_A h_B}-y_{\rm cms}^{\gamma h_{A,B}}$, 
is simply related to the photon momentum fraction via \eqs{\Delta y^{\gamma h_{A,B}}_{h_A h_B} =-\epsilon \frac{1}{2} \ln x_\gamma,}
In order to produce a particle of mass Q at threshold\footnote{\ie\ at rest in the photon-hadron cms.} in a 
photon-hadron collision at $\sqrtsgh$ resulting from a hadron-hadron UPC at $\sqrtsnn$, one approximately has 
$x_\gamma \simeq Q^2 / s_{NN}$. At AFTER@LHC, for $Q^2=4$ GeV$^2$, $x_\gamma= 3 \times 10^{-4}$, which gives 
$\Delta y^{\gamma h_{A,B}}_{h_A h_B} \simeq -4$. This explains the maximum at $y_{\rm cms} \simeq 4$ of the solid 
line of \cf{fig:flux_y_p_on_Pb}.

If the emission is too soft, there is simply 
not enough energy to create a particle of a given mass, $Q$, -- this explains why the curve for $Q^2 =4 (40)$~GeV$^2$ 
drops sharply at $y_{\rm cms}\simeq 4(3)$. In the second case (\cf{fig:flux_y_Pb_on_p}), the flux is the highest in the 
opposite direction where soft photons are emitted by the lead projectile. Since the lead beam
energy is lower ($y_{beam}=4.2$) and the energy cut-off smaller, the particle is less backward than it is forward in 
the first case. In the third case (\cf{fig:flux_y_p_on_p}), both proton can emit. The behaviour is
similar to (a) and (b), but for a lower value due to the absence of the factor $Z^2$. 
It is however harder in the forward (backward) region for a projectile (target) proton emitter -- 
note that the flux in the tail is still nonzero at $y$ down to $\pm 3$.
This  mirrors the  possibility for harder photon (up to $x_\gamma \simeq 0.1$) emission from a proton compared to a larger nucleus.
The solid and dashed curves on \cf{fig:flux_y_p_on_p} refer to two different fluxes: Drees-Zeppenfeld as in $ep$ collisions (solid)
and $b_{\rm min}=2 \times R_p$ (dashed).

The advantage offered by the fixed-target mode is therefore obvious when the emitter is the projectile. In such a case, 
the large rapidity differences between the photon-hadron and hadron-hadron cms and that between the hadron-hadron cms 
and the laboratory frame nearly cancel. This results in the production of the particle
at slightly positive rapidities which are easily covered by typical detectors.

\subsubsection{Double-photon case}

Obviously, one can also consider UPC where both colliding hadrons radiate a photon. This is expected
to be the dominant reaction at work in dilepton production, $AA'\to A \ell^+ \ell^- A'$, PbH$\to$Pb$\ell^+ \ell^-$ H, $p$Pb$\to p \ell^+ \ell^-$ Pb  or
$ pp' \to p\ell^+ \ell^-p'$, via $\gaga \to \ell^+ \ell^-$ (see~\cf{fig:BH}), that is the BH process. 
Combining the fluxes from both hadrons, one can derive
a joint photon flux or $\gaga$ luminosity as function of their invariant mass, $\sqrt{s_\gaga}$ or $W$, and rapidity $Y$,
as it is usually done for $gg$ luminosity at the LHC to discuss $H^0$ production rates, for instance.

\begin{figure}[hbt!]
\begin{center}
\includegraphics[width=0.85\columnwidth]{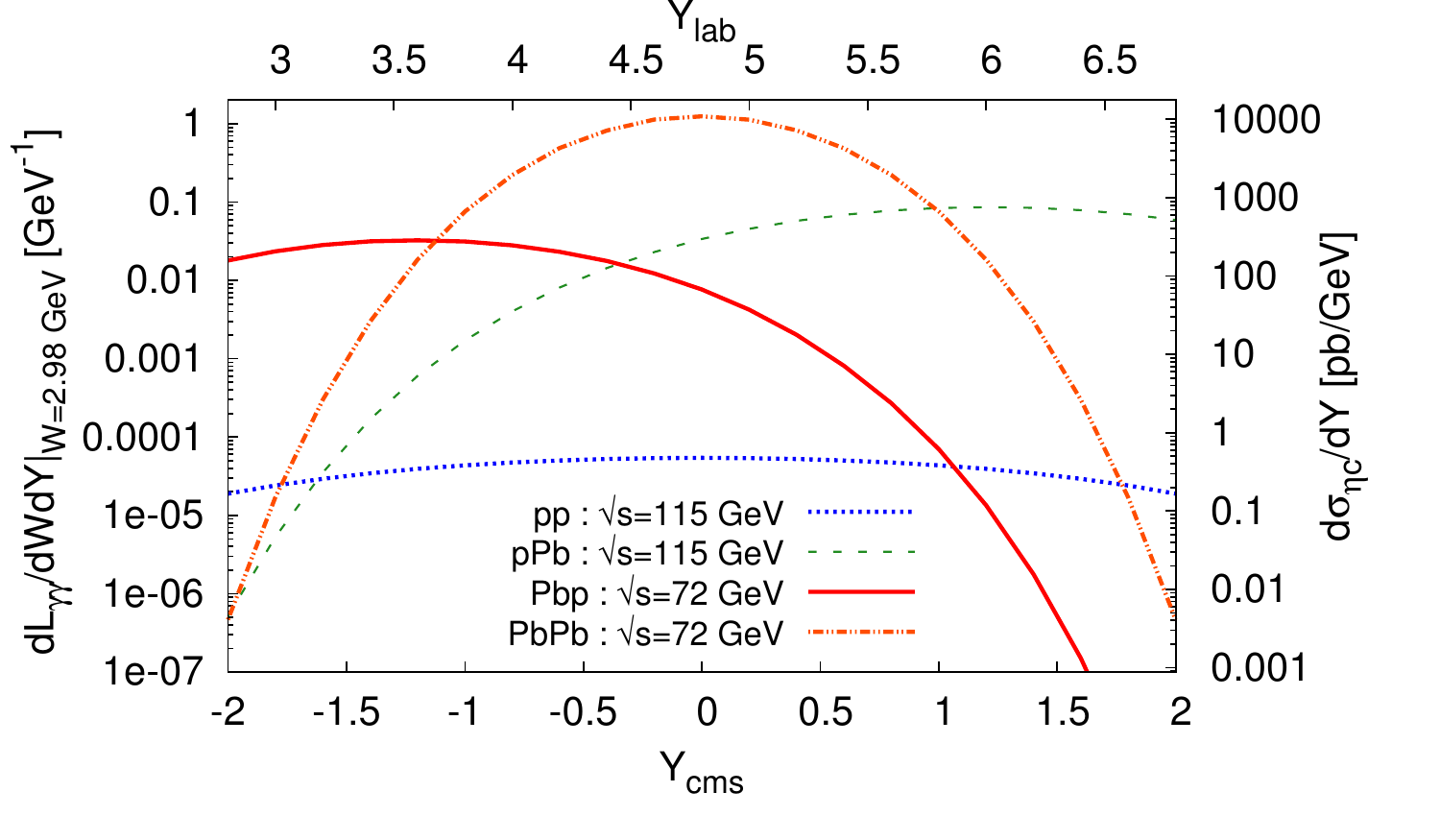}
\caption{$\frac{dL_\gaga}{dW dY}\Big|_{W=m_{\eta_c}}$ (and $\frac{d\sigma_{\eta_c}}{dY}$) as a function of the rapidity, in the hadron-hadron cms (lower $x$-axis) and in the laboratory frame
(upper $x$-axis)\protect\footnotemark.\label{fig:dLdY}}
\end{center}
\end{figure}

\cf{fig:dLdY} shows the rapidity dependence of $\gaga$ luminosity in the hadron-hadron cms  
at a fixed $W$, obtained from  of Eqs. 42, 43, 50, 51 and 52 of 
 \cite{Baur:2001jj} with numerical integrations on the impact parameters. In the case of the BH process 
discussed in detail in the next section, the invariant mass of the pair
is that one the dilepton. One observes that the maxima in the Pb$p$ and $p$Pb fluxes 
occur at respectively 
backward (forward) rapidities because the photon spectrum from the proton is harder than that from the ion.

\footnotetext{The reason why we took  $W=m_{\eta_c}=2.98$~GeV for this example 
will become clear in the next section.}

\section{Lepton-pair production: energy, invariant-mass, rapidity and transverse-momentum dependencies}

We now discuss in more details the dominant process involved in lepton-pair production that is BH
from $\gaga \to \ell^+ \ell^-$ (\cf{fig:BH}) in hadron-hadron collisions.

\begin{figure}[hbt!]
\begin{center}
\subfloat[ BH]{\includegraphics[scale=.35]{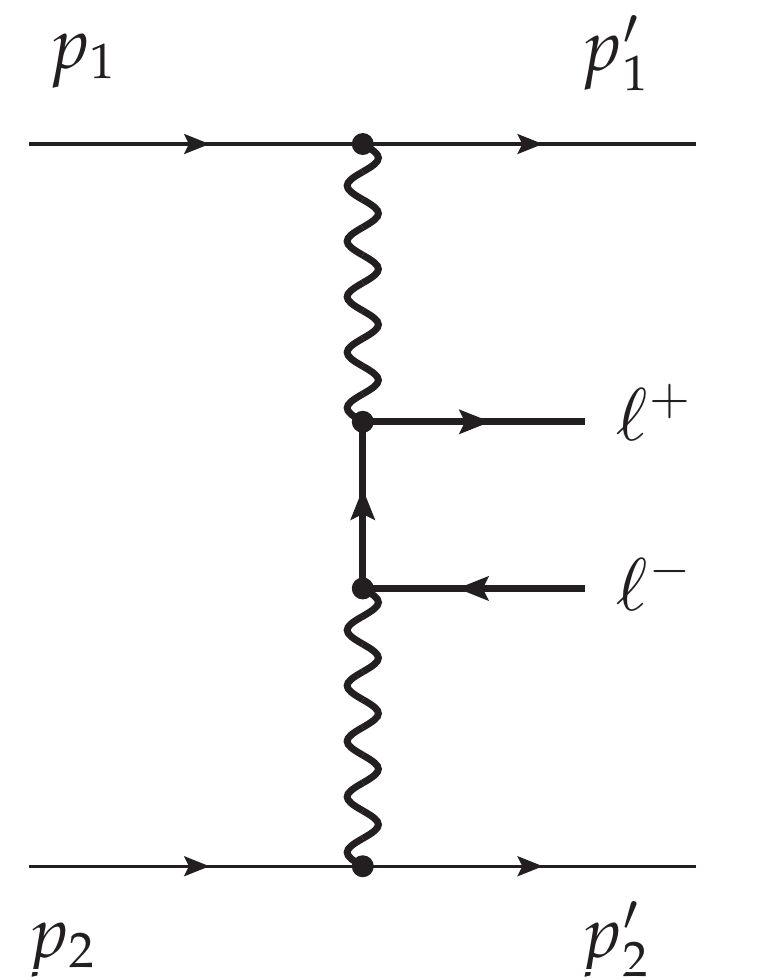}\label{fig:BH}}
\subfloat[ TCS with gluon GPDs]{\includegraphics[scale=.35]{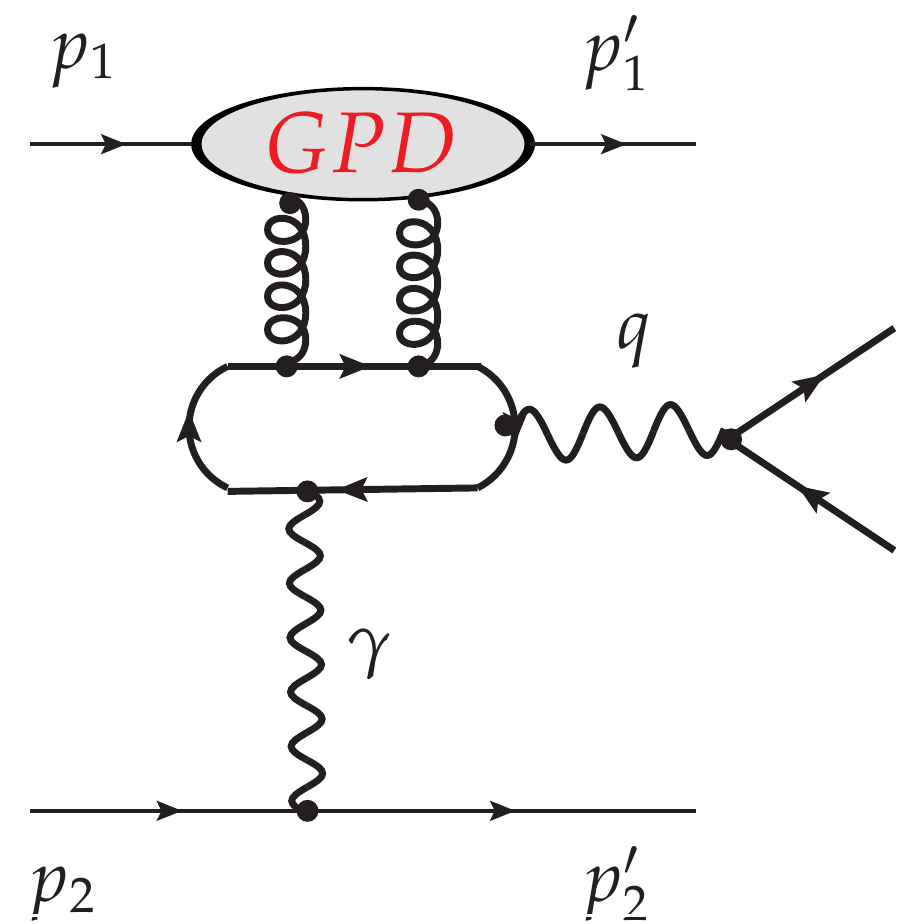}\label{fig:TCS}}
\subfloat[ $\gamma-I\! P$ fusion]{\includegraphics[scale=.35]{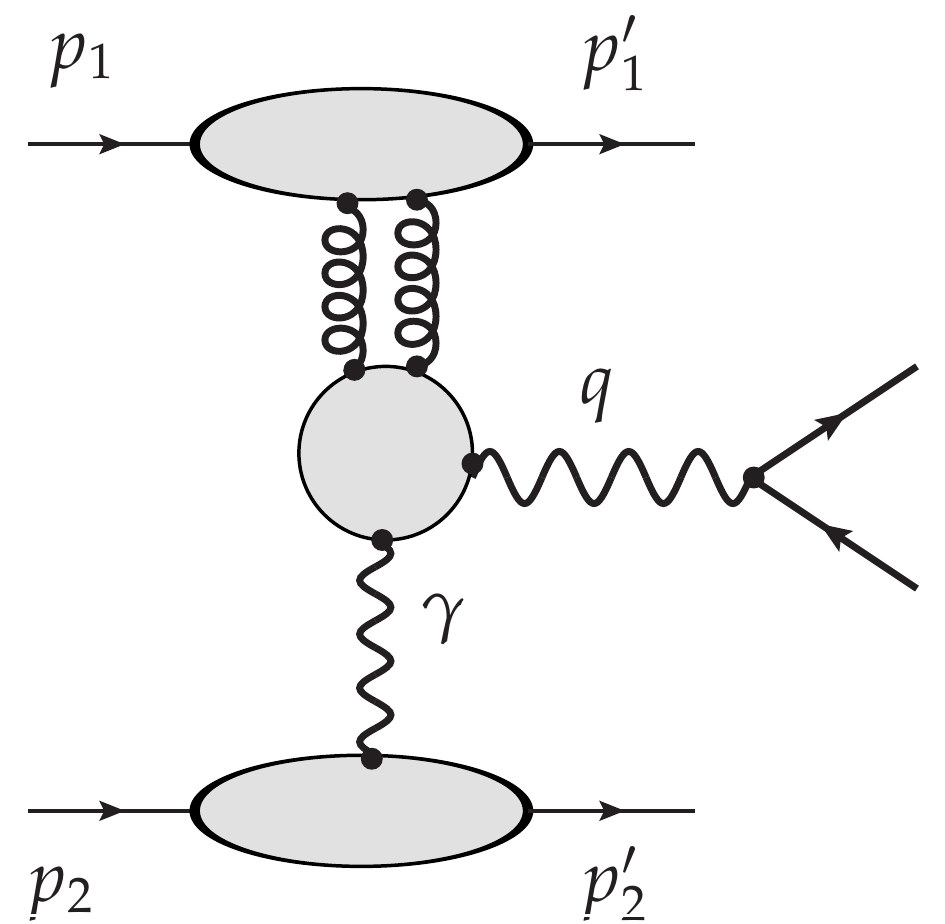}\label{fig:gam-pom-ll}}
\subfloat[ $O-I\!P$ fusion]{\includegraphics[scale=.35]{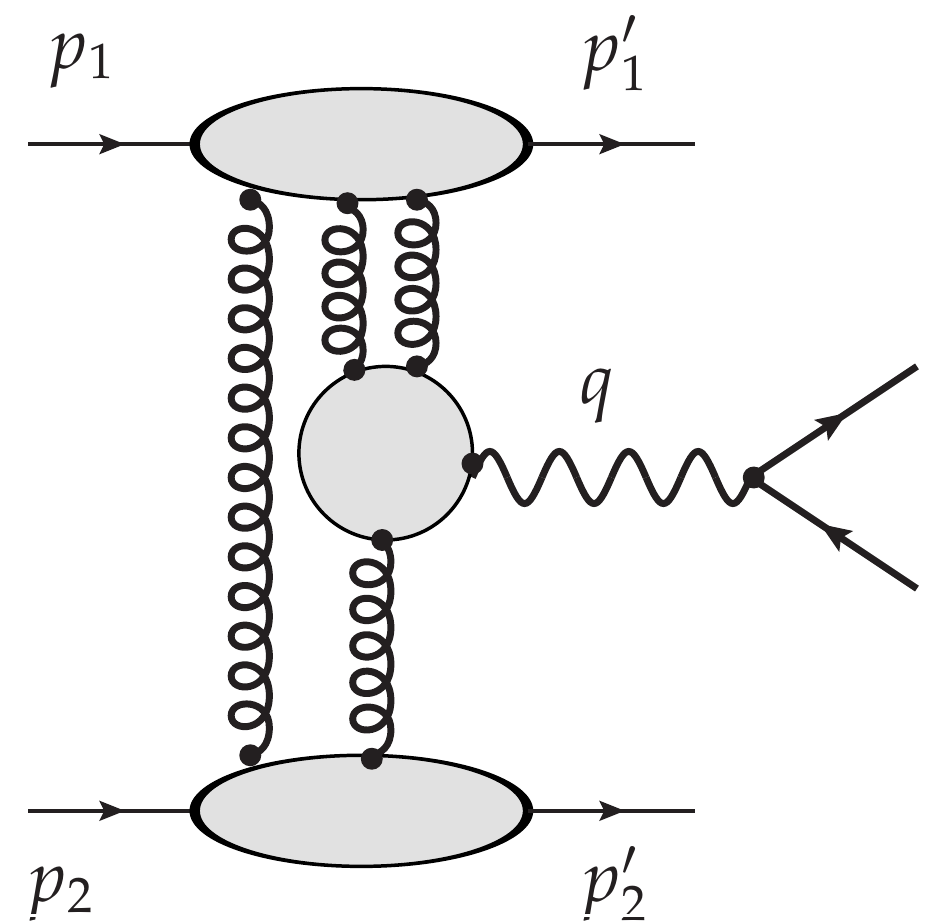}\label{fig:odd-pom-ll}}
\caption{Feynman graphs representing the scattering of two hadrons of momentum $p_1$ and $p_2$ remaining 
intact with final state momenta $p'_1$ and $p'_2$ producing a lepton pair via (a) the BH process 
by photon-photon fusion, (b) via TCS described in terms of a GPD
(here the gluon one), (c) via photon-pomeron fusion represented by the minimal number of gluon exchanges
and (d) via odderon-pomeron fusion also represented with the minimal number of gluon exchanges.}
\label{fig:graphs}
\end{center}
\end{figure}

\subsection{Total cross section and for fixed $Q$}

The total cross section for dimuon production by two heavy and charged particles is well known and can readily be computed 
from the analytical formula of Racah~\cite{Racah:1937zz}. At $\sqrt{s}=115$~GeV, 
one gets 15.0 nb for \pp\ collisions, to be compared with 16.5 nb with the  {\sc Starlight} Monte Carlo 
code\footnote{\href{http://starlight.hepforge.org/}{STARLIGHT website}.}. The corresponding cross section for \pPb\ is simply obtained by multiplying by $Z^2$, \ie\ 100 $\mu$b.
In this case, it is assumed that the particle are point-like. {\sc Starlight} rather gives 36  $\mu$b.

\begin{table}\renewcommand{\arraystretch}{1.4}
\small 
\begin{tabular}{lc|cc|cc|cc|c}
\hline\hline
System & \multicolumn{2}{c}{$pp$} & \multicolumn{2}{c}{$p$Pb} & \multicolumn{2}{c}{Pb$p$} & \multicolumn{2}{c}{PbPb} \\
\hline
$\sqrtsnn$ [GeV] & \multicolumn{2}{c}{115} & \multicolumn{2}{c}{115} & \multicolumn{2}{c}{72} & \multicolumn{2}{c}{72} \\
 & ~BW~ & ~SL~ & ~BW~ & ~SL~ & ~BW~ & ~SL~ & ~BW~ & ~S~~  \\
$\frac{d\sigma_{\ell\ell}(Q=2~{\rm GeV})}{dQ}$ [nb/GeV] & 0.14 & 0.15 & 200 & 210 & 77  & 84  & 7000 & 7100\\
$\frac{d\sigma_{\ell\ell}(Q=2~{\rm GeV}, y_{\rm cms}^{\ell^+ \ell^-}=0)}{dQdy_{\rm cms}^{\ell\ell}}$ [nb/GeV] & 0.039  &0.038 & 39  & 45 & 14  &16  &  5500& 5600 \\
$\frac{d\sigma_{\ell\ell}(Q=2.98~{\rm GeV})}{dQ}$ [nb/GeV] & 0.03 & 0.031 & 32 & 34 & 9.7 & 11  & 230 & 250 \\
$\frac{d\sigma_{\ell\ell}(Q=2.98~{\rm GeV}, y_{\rm cms}^{\ell^+ \ell^-}=0)}{dQdy_{\rm cms}^{\ell\ell}}$ [nb/GeV]~~ & 0.009  & 0.009 & 5.7 & 6.5 &1.3   & 1.6 &200  & 210 \\
\hline\hline
\end{tabular}
\caption{BH differential cross section for fixed dilepton masses integrated (or not) on $y_{\rm cms}^{\ell\ell}$. ``BW'' denotes Breit-Wheeler and ``SL'' denotes {\sc Starlight}\label{tab:BH}.}
\end{table}

\begin{figure}[hbt!]
\begin{center}
\includegraphics[width=0.85\columnwidth]{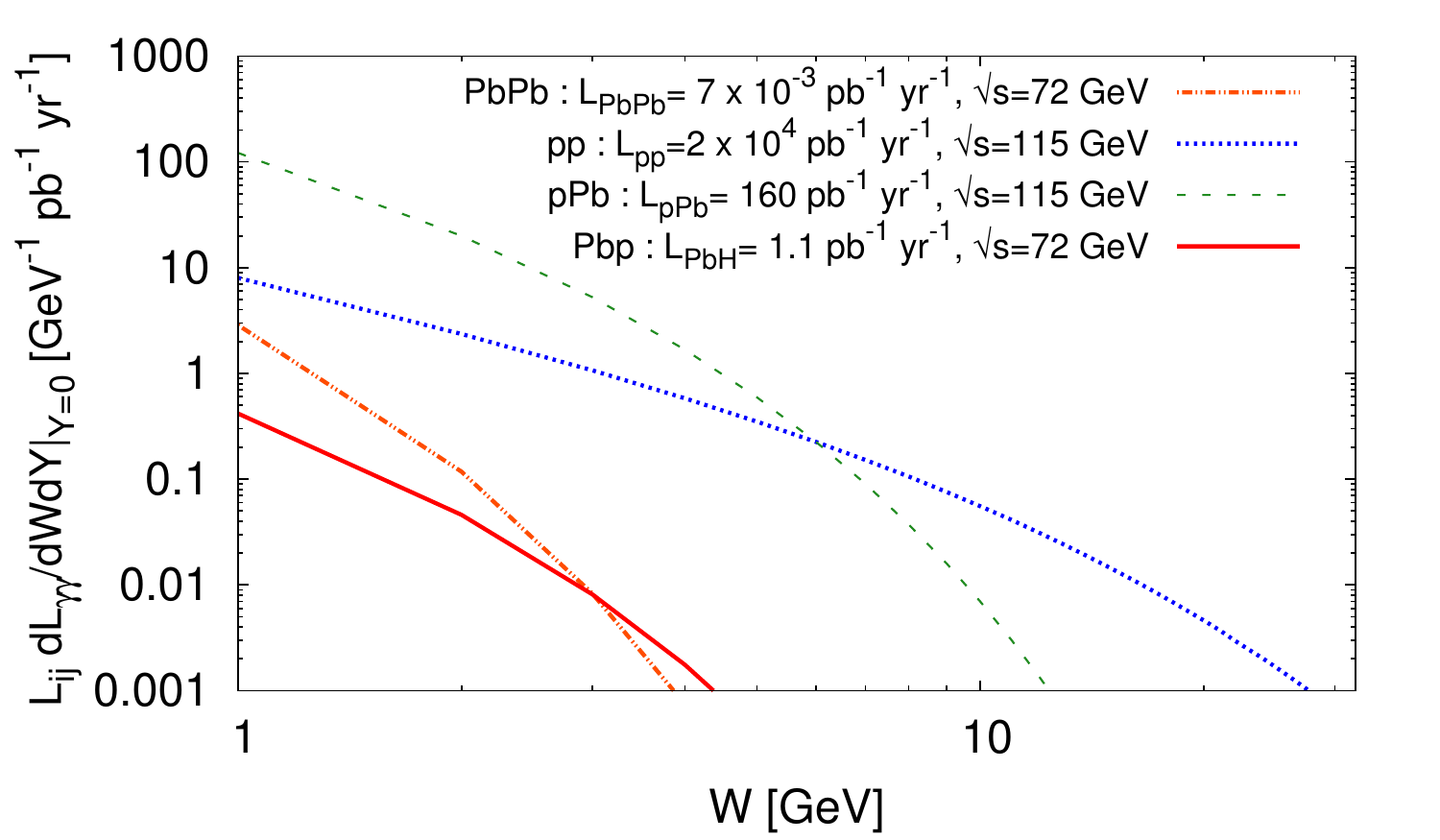}
\caption{${\cal L}_{ij}\times \frac{dL_\gaga}{dW}$ from protons  and from Pb in $pp$, $p$Pb, PbH and PbPb collisions (see \ct{tab:UPC-parameters}) 
as a function of the invariant mass of the photon pair, $W$, at $Y_{\rm cms}$=0 of the $\gaga$ pair (or of the to-be produced dilepton).
\label{fig:dLdWdY}}
\end{center}
\end{figure}

We are however interested in the differential cross sections for
particular values of the dilepton invariant mass, $Q$; muon pairs produced
at $Q \gtrsim2 m_\mu$ are usually difficult to study since the muons have small transverse momenta. 
Such a cross section can easily be obtained by
combining two EPA photon fluxes as done to obtain $\frac{dL_\gaga}{dWdY}$ and then by integrating
over $Y$. One then convolve it with the Breit-Wheeler formula~\cite{Baltz:2009jk}:
\eqs{\label{eq:BW}
\sigma_{\ell\ell}^{\gaga}(Q^2) = {4\pi\alpha^2\over Q^2}
\bigg[&\bigg(2 + {8m_\ell^2\over Q^2}-{16m_\ell^4\over Q^4}\bigg)
\ln{Q+\sqrt{Q^2-4m_\ell^2}\over 2m_\ell} - \sqrt{1-{4m_\ell^2\over Q^2}}
\bigg(1+{4m_\ell^2\over Q^2}\bigg)\bigg].
}

We have checked that we obtained the same result as {\sc Starlight} for dimuon production, 
for instance for $Q=(2,2.98)$ GeV for \pp\ and \pPb\ collisions at $\sqrtsnn=115$~GeV, 
as well as for  \Pbp\ and \PbPb\ collisions at $\sqrtsnn=72$~GeV, 
up to the uncertainty attached to the value taken for 
the nucleus radius, see the first and third rows of results on \ct{tab:BH}.

One readily obtains the rapidity dependence of the differential cross section,
by simply using $\frac{dL_\gaga}{dW dY}\big|_{W=Q,Y=y_{\rm cms}^{\ell\ell}}$ . We already discussed it on \cf{fig:dLdY}.
The $\gaga$ luminosity at a fixed cms rapidity of the photon pair, $Y_{\rm cms}$
(or equivalently $y_{\rm cms}^{\ell\ell}$),
is plotted as a function of $W$  on \cf{fig:dLdWdY} namely at $Y_{\rm cms}=0$ for $pp$, $p$Pb, Pb$p$ and PbPb collisions;
it is multiplied by the corresponding hadron yearly luminosity denoted ${\cal L}_{ij}$.
Just as for the rapidity integrated results, both methods agree for $y_{\rm cms}^{\ell\ell}=0$ as it should be, see the second and fourth rows of results on \ct{tab:BH}.

Knowing $\frac{dL_\gaga}{dW dY}$, one can also obtain the production cross section for a scalar or tensor 
quarkonium, $\cal Q$ provided that we know its partial width into a photon pair, $\Gamma_\gaga$: 
\eqs{\frac{d\sigma^{h_A h_B}_{\cal Q}}{dY_{\cal Q}} \overset{\gaga\to \cal Q}{=} 8 \pi^2 (2J_{\cal Q}+1) \frac{\Gamma_\gaga}{2 M_{\cal Q}^2}\frac{dL_\gaga}
{dW dY}\big|_{W=M_{\cal Q}, Y=Y_{\cal Q}}.}
From $\Gamma_\gaga^{\eta_c}=5.1 \times 10^{-6}$ GeV and $\Gamma_\gaga^{\chi_{c2}}=5.3 \times 10^{-7}$ GeV~\cite{Agashe:2014kda}, one gets
$\sigma_\gaga^{\eta_c}= 8.8$ nb and $\sigma_\gaga^{\chi_{c2}}= 3.2$ nb. 
The rapidity dependence of this cross section is thus up to a constant factor that of the joint flux 
at the corresponding\footnote{Hence our choice of $W=m_{\eta_c}=2.98$ GeV in~\cf{fig:dLdWdY}.}
$W$ (see \cf{fig:dLdY}). In particular, for  \pp\ (\pPb) collisions at $\sqrtsnn=115$~GeV, the $\eta_c$ 
cross sections at $y=0$ in the hadron-hadron cms are 0.5 pb (0.4 nb) and for \Pbp\ (\PbPb) collisions 
at $\sqrtsnn=72$~GeV, 67 pb (11 nb). With the yearly luminosities in \ct{tab:UPC-parameters}, one can respectively 
expect $10^4$, $1.8 \times 10^6$, 74 and 80  $\eta_c$  per year. 

At this point, two remarks are in order. First, we stress that although, 
$m_{\eta_c}$ is above the energy ``cut-off'' showed in \ct{tab:UPC-parameters} 
for the systems \pPb, \Pbp\ and \PbPb, 
the rates are nonzero using realistic photon fluxes. Second, our result for \pp\ collisions 
is 2-3 times below that recently derived by Goncalves and Sauter~\cite{Goncalves:2015hra}. We attribute
this difference to the fact that they used the  Drees and Zeppenfeld photon flux, which is precisely twice larger
than that derived with $b >  2 R_p$ at $x_\gamma \simeq \sqrt{s}/m_{\eta_c} \simeq 0.025$ (see \cf{fig:flux_x}). As we noted above, the former should 
only strictly be used for $ep$ collisions. Such a difference only arises at 'low' hadron-hadron cms energies which correspond to $x_\gamma$ 
on the order of $10^{-2}$ and above. A measurement of the $\eta_c$ or $\chi_{c2}$ cross section in exclusive $pp$ 
collisions at AFTER@LHC is therefore ideal to tell which choice is the most appropriate.

\subsection{Production at nonzero transverse momenta}

If we prefer to look at dileptons produced in reactions characterised 
by a momentum transfer squared of one emitter, $|t|$, up to 1 GeV$^2$, which results in a 
measurable transverse momentum, $q_T$, of the dilepton, it
may be more suitable to treat the photon emission using a nucleon form factor\footnote{In the nuclear case,
such a configuration is admittedly much suppressed.}. By construction, the off-shellness of this photon
cannot be neglected. This amounts to consider the process 
photoproduction, $\gamma p \to \ell^+\ell^- p$, where the beam\footnote{The term ``beam'' may be improper
in the fixed-target case since this photon can very well be emitted by a nucleon or nucleus in the target; this is
particularly true for $p$Pb collisions.}  photon flux is still 
treated in the EPA approximation and the effect of the form factor embedded in the photoproduction
cross section.

\begin{figure}[hbt!]
\includegraphics[scale=.4]{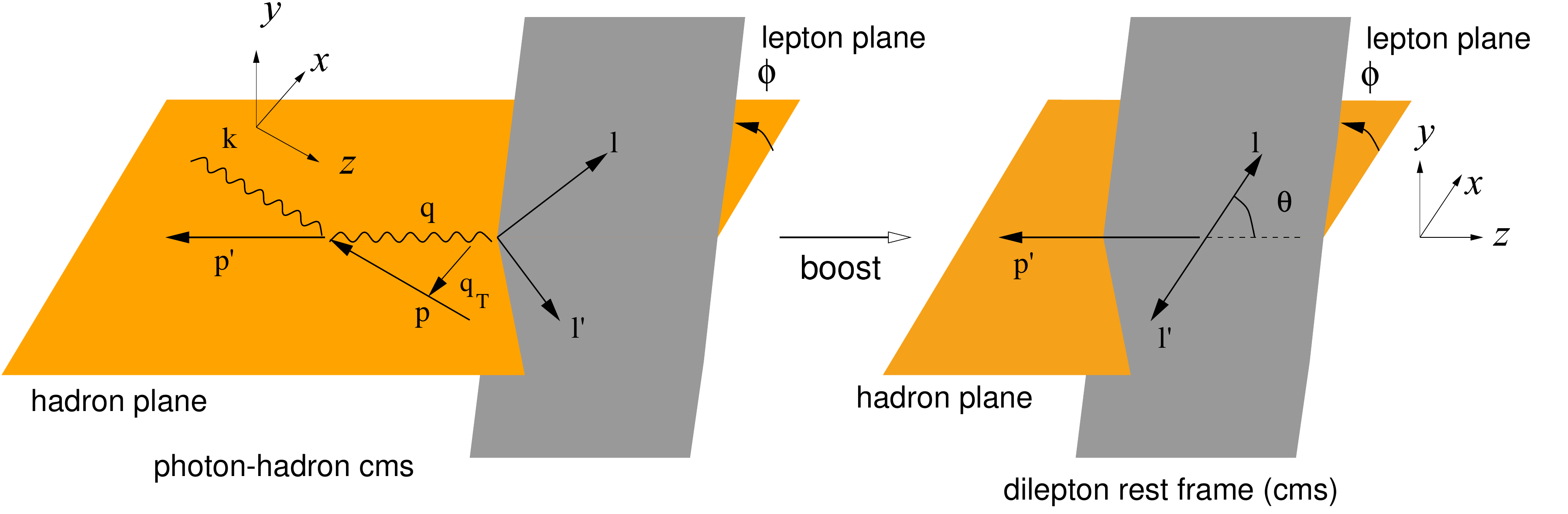}
\caption{Definition of the kinematical variables in $\gamma(k) h(p) \to \ell^+(l')\ell^-(l) p(p')$ in particular the angles $\theta$ and $\phi$ attached
to the $\ell^-$.}
\label{fig:drawing-angles}
\end{figure}

Following \cite{Berger:2001xd},  this $\gamma h$ differential cross section for an unpolarised target to be convoluted 
with the beam photon flux (provided that $\sin{\theta} \gg m_\ell / Q $) reads

\begin{align}
  \label{eq:approx-BH}
&\frac{d \sigma^{\gamma h\, {\rm BH}}_{\ell\ell}}{dQ^2\, \, dq_T^2\, d\cos\theta\, d\phi}(y_{\rm cms}^{\ell\ell})
  \approx  J
    \frac{\alpha^3_{em}}{2\pi s_{\gamma h}^2}\, \frac{1}{-t}\,
    \frac{1 + \cos^2\theta}{\sin^2\theta} \, 
  \\ &\times \left.\left[ \Big(F_1^2(t) -\frac{t}{4M_N^2} F_2^2(t)\Big)
            \frac{2 (s_{\gamma h}-M_N^2)^2}{Q^4} \frac{q_T^2}{-t}
        + (F_1(t)+F_2(t))^2 \,\right]\right|_{t=t(y_{\rm cms}^{\ell\ell},q^2_T,Q,\epsilon)},\nonumber
\end{align}
where $F_{1,2}$ are respectively the Dirac and Pauli form factors evaluated at $t=(p-p')^2=(k-q)^2$, 
the quantity $\epsilon$ and procedure to obtain 
$t(y_{\rm cms}^{\ell\ell},q^2_T,Q,\epsilon)$ and $J$ (the Jacobian to change from $t$ to $q^2_T$) are  explained
in the appendix \ref{appendix-1}. The $\theta$ and $\phi$ angles are defined as on \cf{fig:drawing-angles}.
The apparent divergence at $\theta\to 0$ is regulated when the lepton mass is kept, just as in the Breit-Wheeler
equation, \ce{eq:BW}, which is logarithmically divergent for $m_\ell\to 0$. We note that 
$\theta$ -- the polar angle of the lepton -- which is
defined in the rest frame of the dilepton, can only be approximately related to the (pseudo)-rapidity of 
the lepton in the cms frame for $y_{\rm cms}^{(\ell\ell)}\simeq 0$ and $q_T \ll Q$. In such a case, the particular configuration
$\theta \to \pi/2$ corresponds  to $y_{\rm cms}^\ell \simeq \eta_{\rm cms}^\ell \simeq 0$ which falls into the acceptance of a detector like LHCb and ALICE (with
Pb beams).
Using \ce{eq:rap-dep-dndy}, one then obtains $d \sigma^{hh\, {\rm BH}}_{\ell\ell}/(dQ^2 \, dq_T^2\, d\cos\theta\, d\phi\, dy_{\rm cms}^{\ell\ell})$.

\section{Timelike compton scattering  or exclusive photoproduction of a dilepton}

The process of the lepton-pair production in ultraperipheral collisions may also be used to investigate 
hadron structure in terms of GPDs through the measurement of the contribution of the TCS process, \cf{fig:TCS}, to the cross section. 
Although the BH amplitude squared is much larger than the TCS one, 
it is possible to study the interference term between TCS and BH processes, which may be projected out through the
analysis of the angular distribution of the produced leptons \cite{Berger:2001xd} and which depends on the GPDs. 
The interference term of the differential cross section is given by:

\begin{align}
\label{eq:approx-BH-INT}
\frac{d \sigma^{\gamma h\, {\rm INT}}_{\ell\ell}}
{d Q^2\, dt\, d\cos\theta\, d\phi}
&\approx -
\frac{\alpha^3_{em}}{4\pi s_{\gamma h}^2}\, \frac{\sqrt{t_0-t}}{-tQ}\,
\frac{\sqrt{1-\eta^2}}{\eta}\,\left( \cos\phi\, \frac{1+\cos^2\theta}{\sin\theta}\right) \\
&
\times \re\left[ F_1(t) {\cal H}(\eta,t) - \eta (F_1(t)+F_2(t))\, \tilde{\cal H}(\eta,t) -
\frac{t}{4M^2} \, F_2(t)\, {\cal E}(\eta,t) \,\right] \,,\nonumber
\end{align}
where we have neglected lepton mass and assumed that $s_{\gamma h},Q^2 \gg t, M_N^2$. The variable $\eta$ called {\it skewedness} is given by:
\begin{equation}
\eta = \frac{Q^2}{2s_{\gamma h}-Q^2},
\end{equation}
and $t_0$ is maximal value of squared momentum transfer $t$ reached at $q_T=0$, and is equal $t_0 = -4M^2{\eta^2 \over 1-\eta^2}$
up to corrections in $1/Q^2$. The functions ${\cal H}(\eta,t)$, $\tilde{\cal H}(\eta,t)$ and ${\cal E}(\eta,t)$ 
are the well known Compton form factors. These involve the GPD $H$, $\tilde H$ and $E$ (as defined in~\cite{Diehl:2003ny}) 
respectively through a convolution with the hard-scattering kernels 
$T^{q,g}_{\{H,\tilde{H},E\}}$ computed at a given order in $\alpha_s$:
\begin{align}
&\{\mathcal{H}, \tilde{\mathcal{H}}, \mathcal{E} \} (\eta,t)= \\ &\int_{-1}^1dx \,
\Big[\sum_q  T^q_{\{H,\tilde{H},E\}}(x,\eta)\, \{H^q,\tilde{H}^q,E^q \} (x,\eta,t)+ 
T^g_{\{H,\tilde{H},E\}}(x,\eta)\{H^g,\tilde{H}^g,E^g \} (x,\eta,t) \Big].\nonumber
\end{align}
The expression for the kernels $T_{\{H,\tilde{H},E\}}$ at LO are given in the appendix \ref{appendix-2}.
The Next-to-Leading Order (NLO) hard-scattering kernel\footnote{As done in \cite{Pire:2011st}, we
set $\mu_R=\mu_F=Q$ in the hard-scattering kernel and, as for other similar phenomenological analyses,
 the GPDs are not evolved. In the case of the G-MSTW model, the GPDs can be made $\mu_F$ dependent through the evolution of the input PDFs.}, 
which we use here, can be found in~\cite{Pire:2011st,Muller:2012yq}.

To advocate that the TCS measurement is feasible in the  AFTER kinematics, we present phenomenological 
predictions making use of two GPD models through  double distributions~\cite{Radyushkin:1997ki}: the 
first is based on the Goloskokov-Kroll (GK) model based on fits of meson electroproduction data  and 
the second is a model using MSTW8 parton distribution function with a simple factorised $t$ dependence, referred to as G-MSTW  
(for a detailed description see \cite{Moutarde:2013qs}).

\begin{figure}[hbt!]
\begin{center}
 \subfloat{\includegraphics[width= 0.48\textwidth]{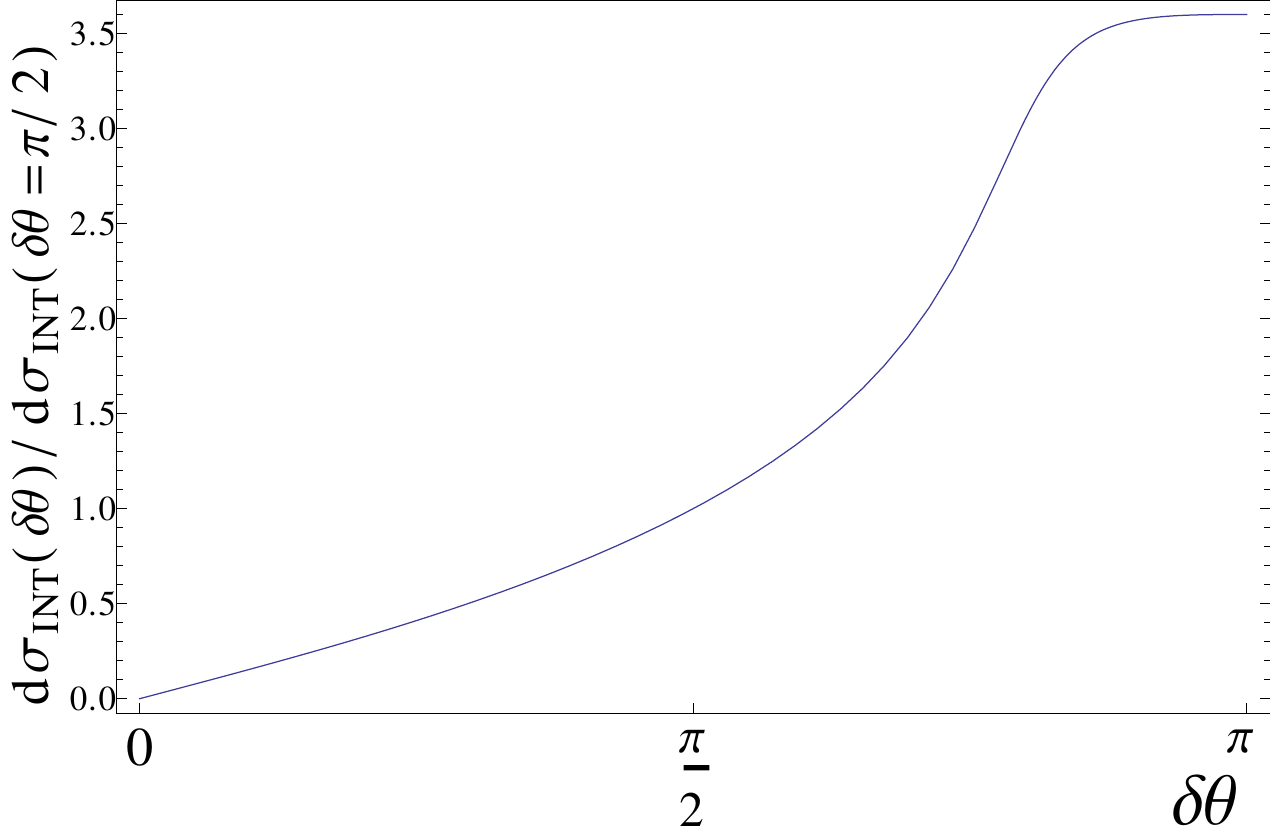}} \ 
 \subfloat{\includegraphics[width= 0.48\textwidth]{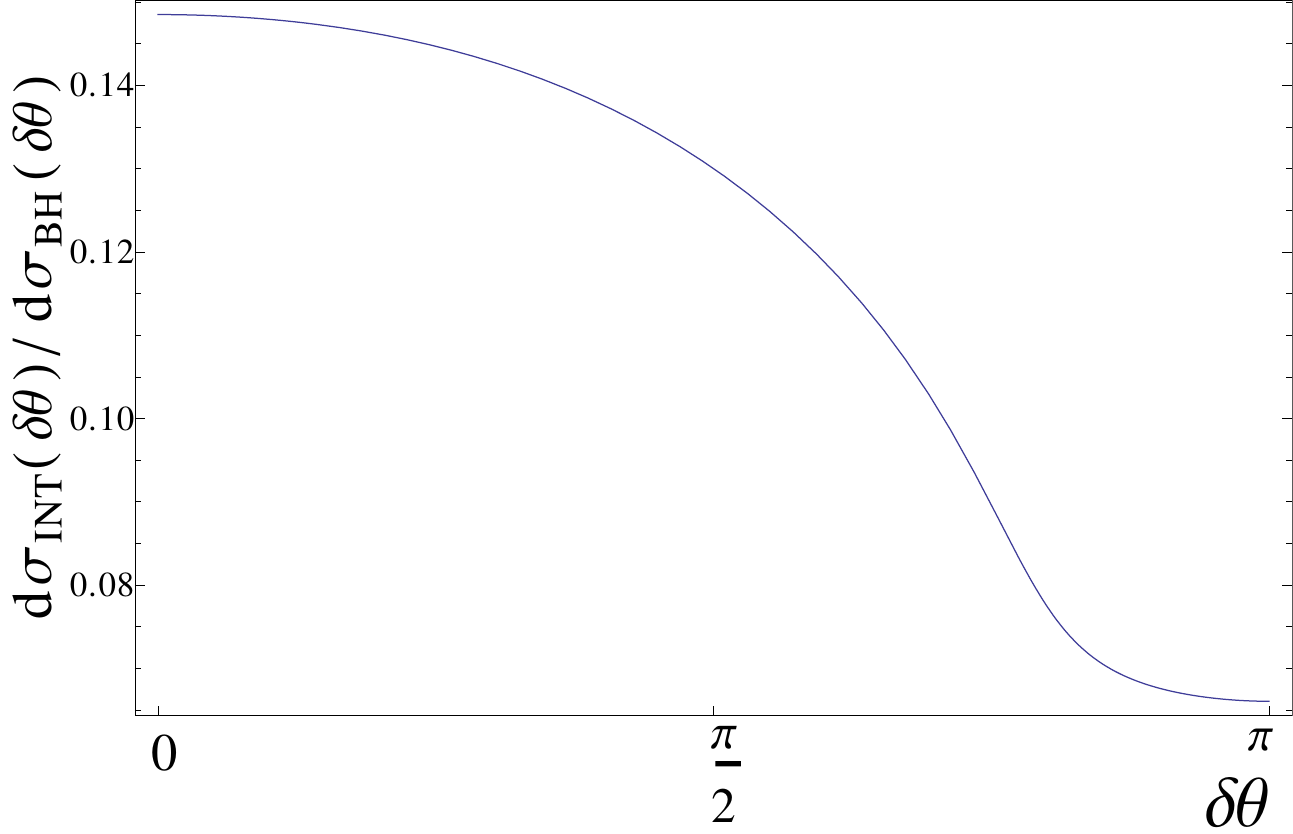}} 
\caption{(left)  $\frac{d \sigma_{\rm INT}}{dy}(\delta \theta)/\frac{d \sigma_{\rm INT}}{dy}(\delta \theta=\pi/2)$, 
(right) $\frac{d \sigma_{\rm INT}}{dy}/\frac{d \sigma_{\rm BH}}{dy}(\delta \theta)$ for the model KG for 
$Q^2 = 4 \gev^2$, $t=-0.1 \gev^2$, and $\phi =0$ for proton run on Pb, integrated over  $\theta \in (\pi/4,3\pi/4)$ and for $y=0$ at NLO.}
\label{fig:ratio Int to BH delta theta}
\end{center}
\end{figure}

Comparing the $\phi$ dependence in \ce{eq:approx-BH} to that in \ce{eq:approx-BH-INT}, one directly sees that one can 
project out the interference term and eliminate the unwanted BH background by integrating the 
differential cross section over $\phi$ with the weight $\cos\phi$. To avoid the limit $\theta \to 0$ where the BH 
signal gets very large, one can integrate over $\theta$ symmetrically around $\pi/2$ up to a value which depends
on the size of the interference. Varying the limits of integration in $\theta$ according to
\begin{equation}
\theta \in \left(\frac{\pi- \delta \theta}{2} , \frac{\pi+ \delta \theta }{2}\right),
\end{equation}
results in the change of the strength of the interference BH-TCS with respect to the BH cross section shown on 
\cf{fig:ratio Int to BH delta theta}. On the left panel, we show the ratio 
of the the integral of the interference BH-TCS for a given $\delta \theta$ normalised to that for $\delta \theta = \pi/2$. 
We see that for $\delta \theta$ close to $\pi$ the ratio of the integrals of the interference stops increasing. 
In any case, in this limit, as we present on the right panel, the ratio of the integral of the interference term to the BH one
gets very small because the BH cross section gets large\footnote{In one neglects the lepton, it even diverges.}.
In other words, there is no specific reason to take $\delta\theta$ much larger than $\pi/2$; below this value the
magnitude of the interference term is close to 10 \% of the BH term.

One can then combine the $\gamma h$ cross section with the photon flux as a function of the dilepton rapidity at a fixed
$t$ to get the interference part of 5-fold  hadron-hadron differential cross section. Defining the integration region as
\eqs{\int_\Omega d^5F\equiv \int_{\pi/4}^{3\pi/4} \!\!\!\!d \theta \int_0^{2\pi} \!\!\!\!d \phi \int_{-2.5}^0 \!\!\!\!d y \int_{0}^{0.25} \!\!\!\!dq^2_T \int_{1.5}^3 dQ,} 
the BH cross section from a 7 TeV proton beam on a Pb target is
\begin{equation}
\sigma^{\text{p} Pb}_{\rm BH}=\int_\Omega d^5F  
\frac{d \sigma_{\rm BH}}{dQ dq^2_T dy d\theta d\phi} = 1940 ~\textrm{pb},
\end{equation}
which, for a luminosity of 0.16 fb$^{-1}$yr $^{-1}$, gives about $3 \cdot 10^5$ events per year. In the case of a H target, one has
\begin{equation}
\sigma^{\text{pH}}_{\rm BH}=\int_\Omega d^5F  
\frac{d \sigma_{\rm BH}}{dQ dq^2_T dy d\theta d\phi} = 7.1 ~\textrm{pb},
\end{equation}
which, for a luminosity = 20 fb$^{-1}$ yr$^{-1}$, gives  $1.4\cdot 10^5$ events per year for a 100 cm liquid-hydrogen target.
Finally, for Pb on H, one has
\begin{equation}
\sigma^{\rm Pb H}_{\rm BH}= \int_\Omega d^5F 
\frac{d \sigma_{\rm BH}}{dQ dq^2_T dy d\theta d\phi} = 5500 ~\textrm{pb},
\end{equation}
which, for a luminosity = 11 nb$^{-1}$ yr$^{-1}$, gives  $6\cdot 10^3$ events per year for a 100 cm liquid-hydrogen target. As aforementioned, 
we do not consider the case where the nucleus is emitter with a $|t|$ of a few hundred MeV$^2$ which should be treated
with nucleus form factors and nuclear GPDs. With a magnitude of 10 \% for the interference term, the azimuthal modulation should be
observable in the 3 cases.

On the \cf{fig:cs and ratio p on Pb}, we show the magnitude of the 3 terms as function of the rapidity as well as the relative
magnitude of the interference term with respect to the BH one for three cases of collisions (a) proton beam on a Pb target, (b) 
Pb beam on a H target, (c) proton beam on a H target. On the left panels, we present the rapidity dependence of the
 BH, TCS and BH-TCS terms of differential cross sections $\frac{d\sigma}{dy dQ^2 dt d\phi }$ evaluated at $Q^2 = 4 \gev^2$,  $t=-0.35\gev^2$ and $\phi =0$
and integrated over $\theta \in (\pi/4,3\pi/4)$ for the GK model. On the right panels, 
we show the ratio of the interference signal to the BH for the GK and 
G-MSTW model. We see that for all cases the signal to the background ratio is around 10-15\%.

\begin{figure}[hbt!]
 \begin{center}
\subfloat[~7 TeV $p$ on Pb target]{\includegraphics[width= 0.5\textwidth]{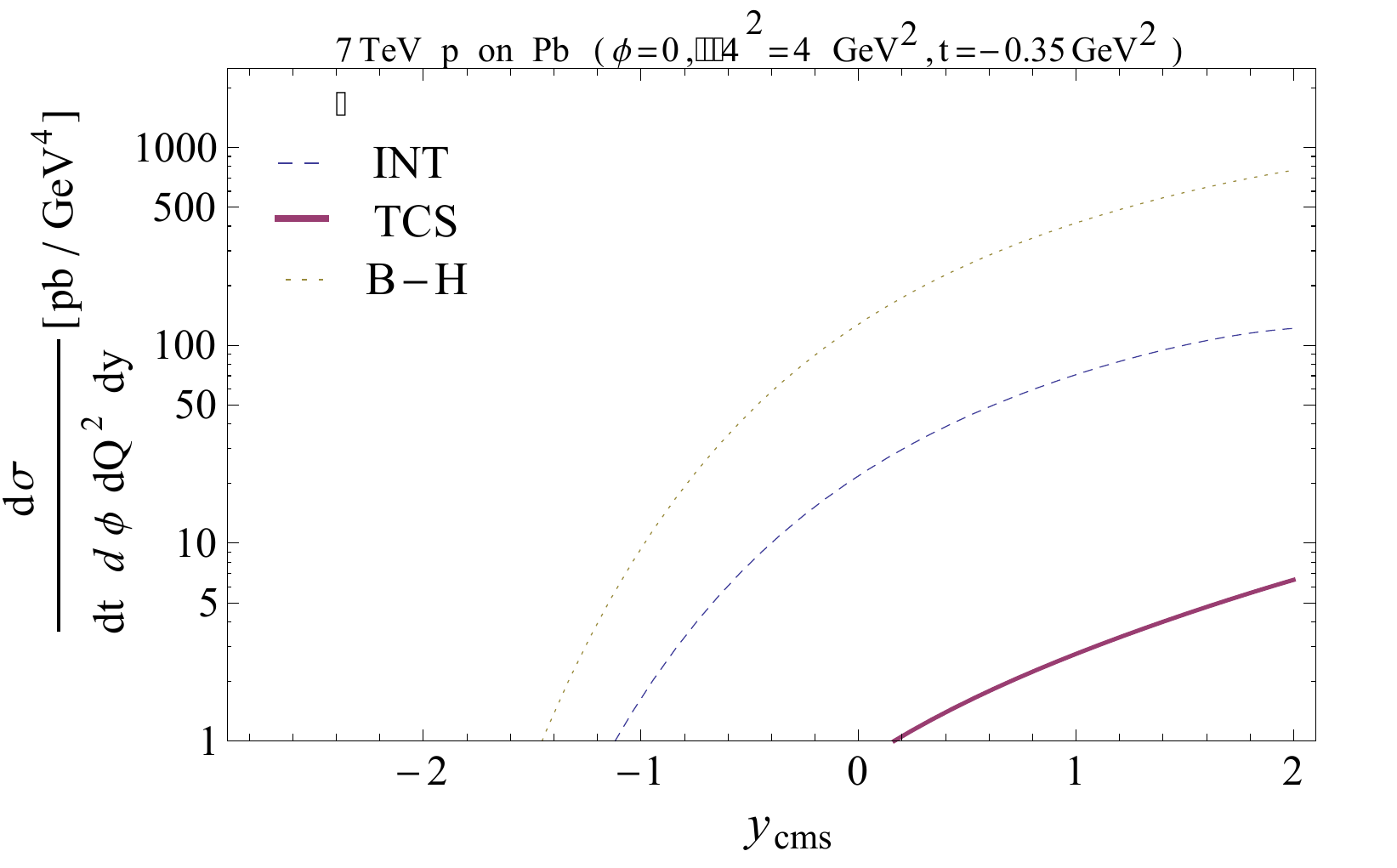}
  \includegraphics[width= 0.5\textwidth]{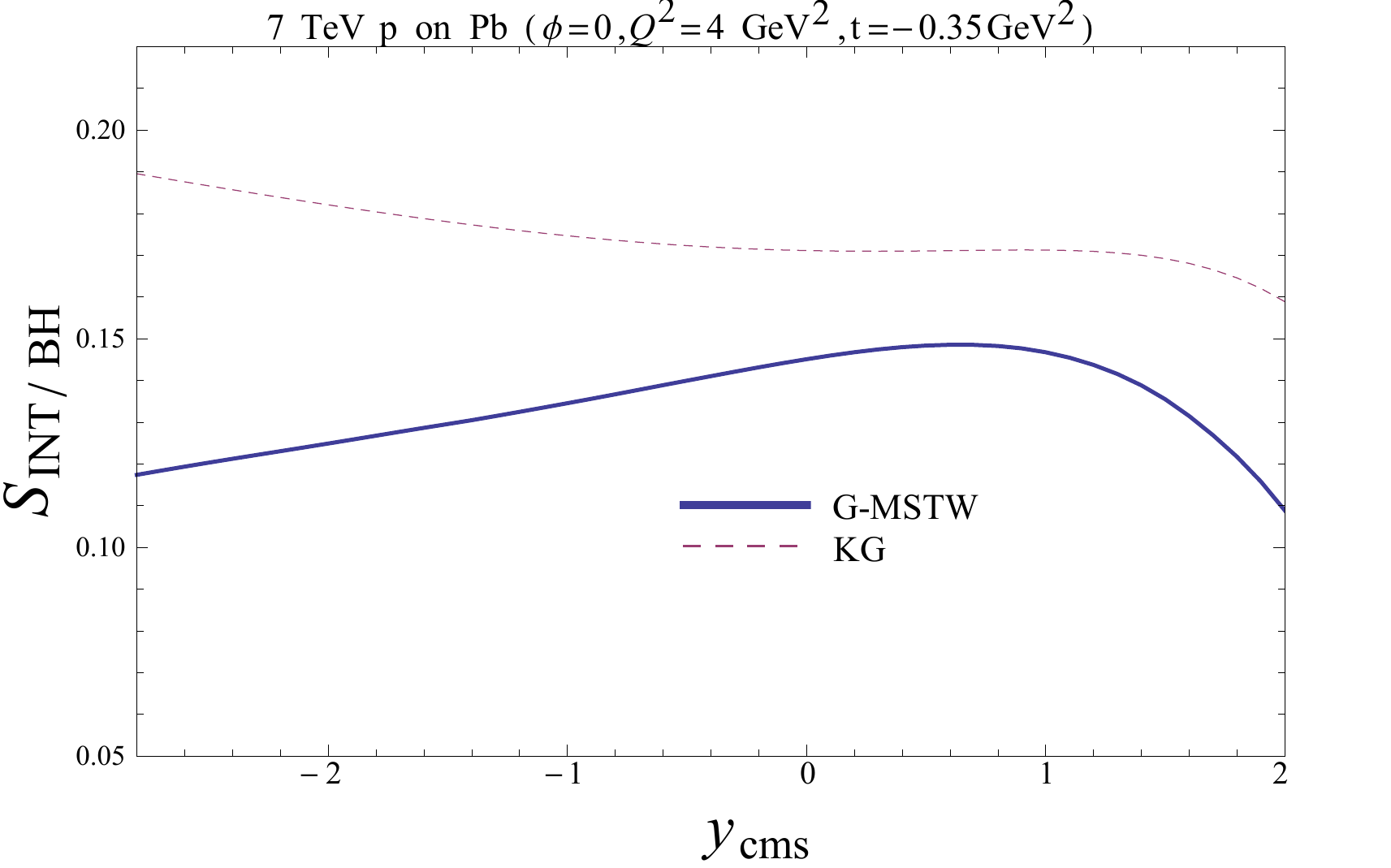}} \\
\subfloat[~2.76 TeV Pb on H target]{  \includegraphics[width= 0.5\textwidth]{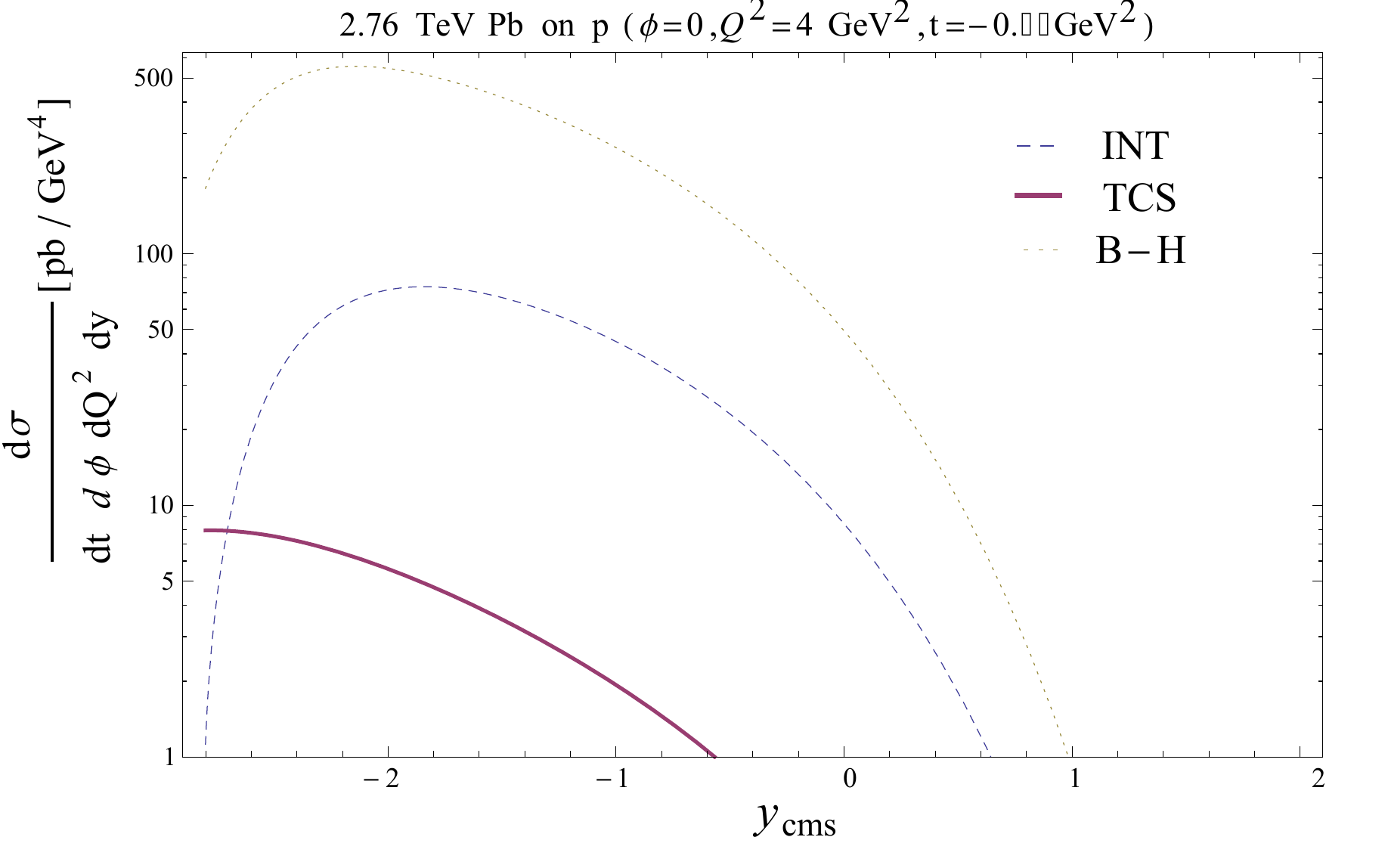} 
  \includegraphics[width= 0.5\textwidth]{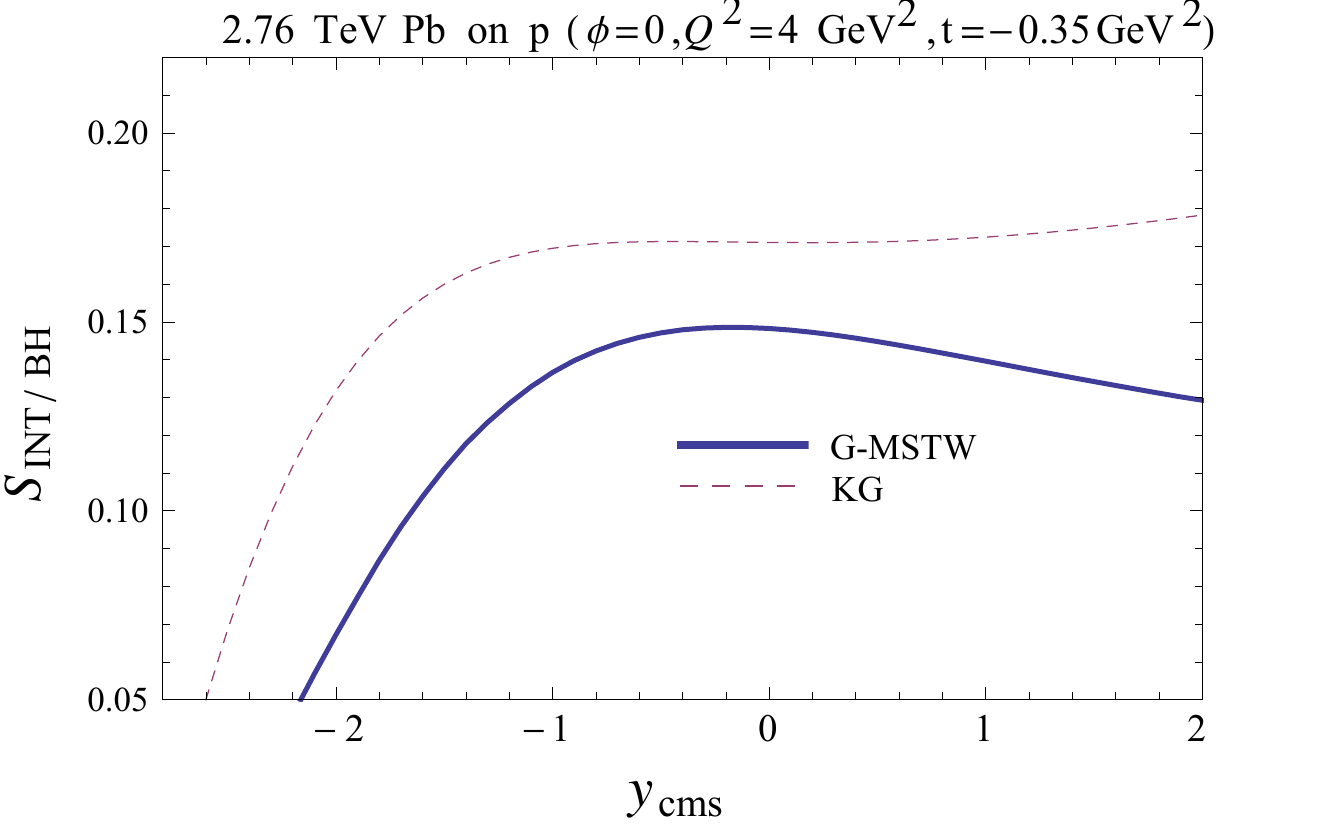}}\\
\subfloat[~7 TeV $p$ on H target]{  \includegraphics[width= 0.5\textwidth]{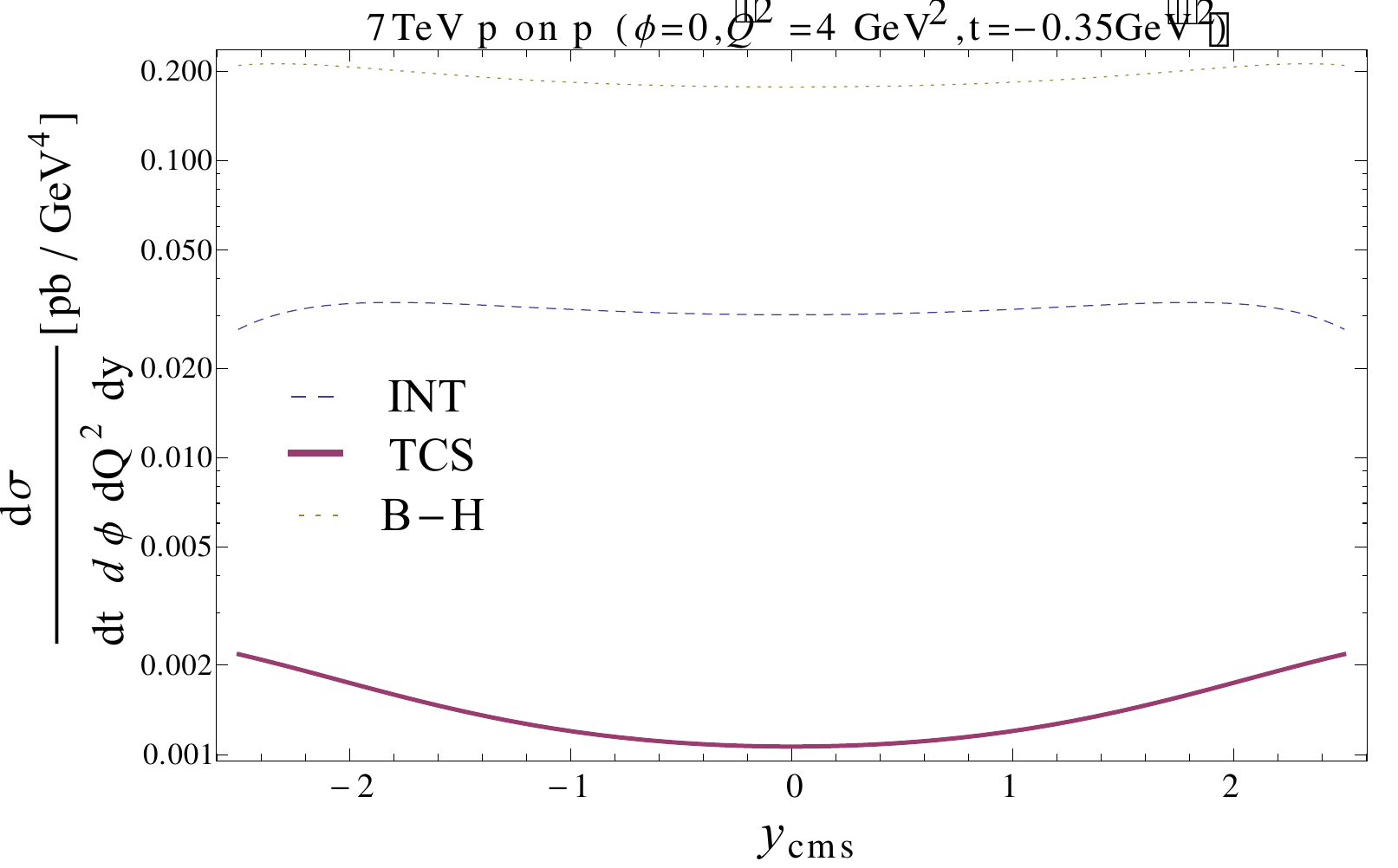} 
  \includegraphics[width= 0.5\textwidth]{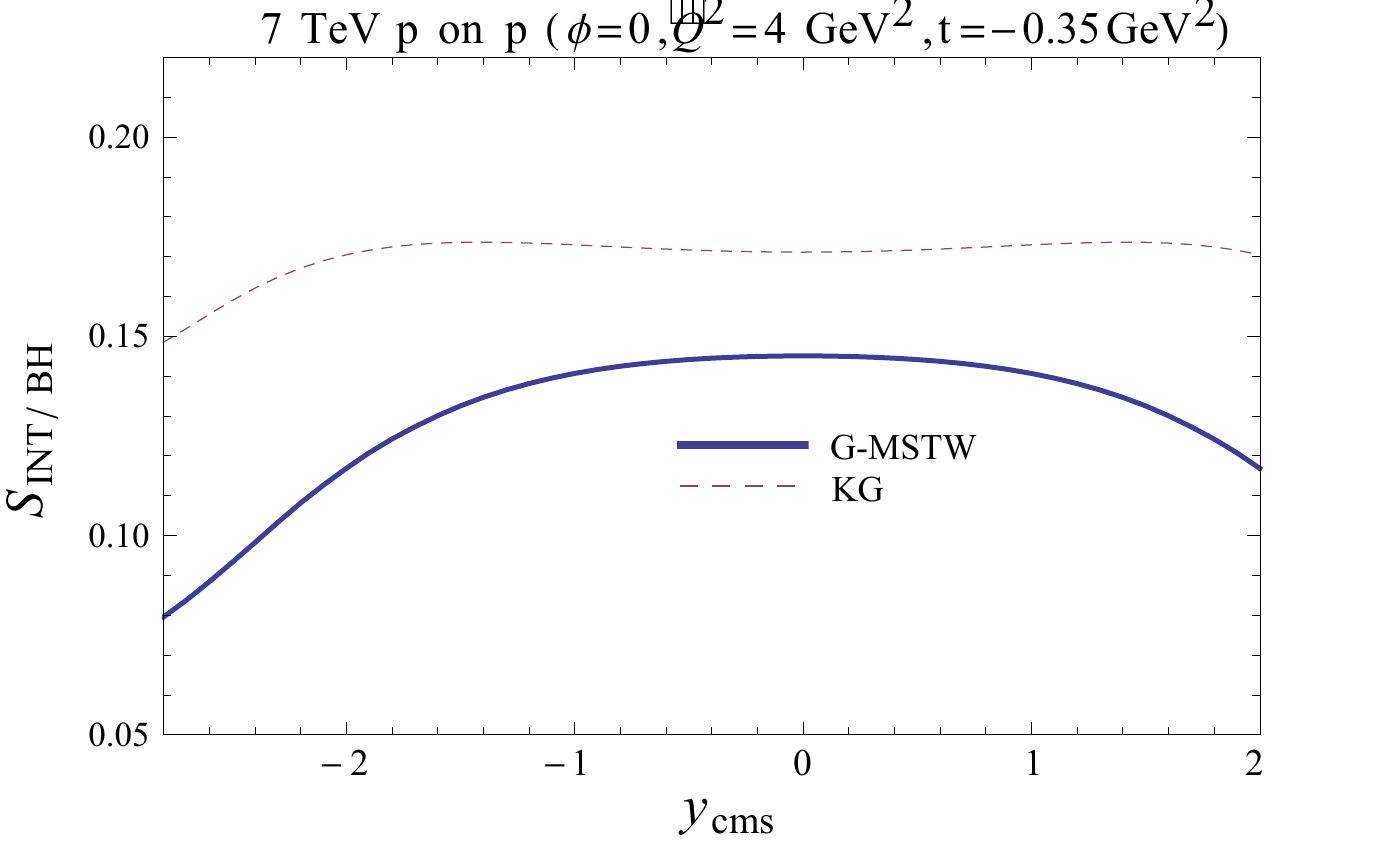}}
\caption{(Left) Differential cross section $\frac{d \sigma}{dy dt d \phi dQ^2}$ for the KG model for $Q^2 = 4 \gev^2$, $t=-0.35 \gev^2$ and $\phi =0$ integrated over  $\theta \in (\pi/4,3\pi/4)$. Dotted line : B-H, dashed line : interference term, solid line: TCS. (Right) Ratio of the interference to BH differential cross section $\frac{d \sigma}{dy dt d \phi dQ^2}$ calculated for $Q^2 = 4 \gev^2$, $t=-0.35 \gev^2$ and $\phi =0$ for the GK (dashed) and G-MSTW (solid) models at NLO.From (a) to (c) $p$Pb, PbH and $p$H cases.}
\label{fig:cs and ratio p on Pb}
\end{center}
\end{figure}

\section{Exclusive lepton-pair hadroproduction via odderon-pomeron fusion}\label{sec:odderon}

Throughout this work, we have based our discussion on the idea that the
 (theoretical and experimental) requirements for selecting UPC collisions
was {\it de facto} preventing hadronic interactions to take place or
at least that photon-induced processes would be dominant. This is a rather
safe assumption in nucleus-nucleus collisions where the coherent photon fluxes
are enhanced by a factor $Z^2$ for each nucleus, whereas double-pomeron-induced reactions 
only scale like $A^{1/3}$. It also seems a reasonable assumption for $pA$ collisions that
electromagnetic interactions dominate.

In the case of $pp$ collisions, the $Z^2$ factor is absent and 
the UPC requirements are also very similar, if not
identical, in practice, to that for an exclusive or diffractive process. For instance, 
in order to impose $b > 1 /R_p$, one can impose that the proton expected to emit the photon is
 only deflected by a $\Delta p_t$ of say maximum 100 MeV, 
which corresponds to $b\gtrsim 2$~fm. Such a requirement however may not
be sufficient to prevent any scattering by the exchange of a pomeron or an odderon.
In fact, the photon-pomeron process depicted on \cf{fig:gam-pom-ll} should in principle
be encompassed in the GPD description at high energies.

The production of a lepton pair in UPC which we discussed so far should
thus be confronted to a potentially competing process leading to the same final
state, in which the virtual photon decaying into lepton pair is produced 
in an Odderon-Pomeron (OP) fusion as depicted in \cf{fig:odd-pom-ll}. This is similar to the situation occurring 
in the case of the central exclusive hadroproduction of a $J/\psi$ 
studied in \cite{Bzdak:2007cz}. In this case, the $J/\psi$  can be produced either by 
the photon-Pomeron fusion -- that is the one expected for UPC -- 
or by the OP fusion. In fact, the cross sections for both processes
are similar at the Tevatron. Owing to the increase of
the gap survival probability for lower energies, it is reasonable to believe that, at
AFTER@LHC energies, \ie~around $\sqrt{s_{NN}}=115$ GeV, exclusive $J/\psi$ production in $pp$ collisions would be dominated by
OP fusion.

Both production mechanisms are however characterized by a different 
$J/\psi$  transverse-momentum dependence~\cite{Bzdak:2007cz}. In particular, if one imposes that
the final state protons have $|t| > 0.25$ GeV$^2$, the OP contribution can be enhanced 
ten times with respect to the photon induced one.

In the case of lepton-pair production in the conditions of UPC,  
a similar situation will likely occur. The experimental study of such a process
with specific cuts on $t$ may thus provide an unexpected path to the Odderon discovery via
interference with the pure QED BH process in the same way as the TCS signal 
would be extracted in the region where it dominates. 
Despite its interest, the evaluation of the corresponding cross section 
along the line of~\cite{Bzdak:2007cz} is far beyond the scope of the present work.

Similarly, the odderon could also contribute to the $\eta_c$ production as recently discussed
in~\cite{Goncalves:2015hra}. $\eta_c$  can be produced by photon-odderon fusion due its different
quantum number with respect to the $J/\psi$. In the $pp$ case, the cross section we obtained 
from $\gaga$ fusion is slightly larger that from photon-odderon fusion obtained 
in~\cite{Goncalves:2015hra}, both on the order of a picobarn or less. 
In the $p$Pb and Pb$p$ cases, the photon can be radiated by the ion and the pomeron by
the proton. As we noted, although $m_{\eta_c}$ is in principle above the so-called energy ``cut-off'', the
photon fluxes, even accounting from the minimum impact parameter for a proton-lead UPC,
 is not zero. In the Pb$p$ case, the photon-odderon induced cross section ranges from 30 up to 360 pb
whereas the $\gaga$ induced one is on the order of 440 pb. As for the dilepton case, 
a study of the transverse-momentum dependence should be able to discriminate between both processes.

\section{Conclusion} 

We have theoretically investigated the feasibility  of accessing the lepton-pair
production in ultraperipheral collisions  at the proposed fixed-target 
experiment AFTER@LHC, which takes advantages of the multi-TeV proton and ion
beams of the LHC. To this aim, we have first estimated the magnitude of the cross
section for lepton-pair production from the fusion of two quasi-real
photons emitted by the quasi-grazing hadrons. This purely electromagnetic BH 
process can serve as an important tool for the determination of the
luminosities  with nucleon or ion beams but  it can also be used for an experimental 
verification of the validity of the effective-photon approximation usually applied 
to estimate the flux of quasi-real photons emitted by these relativistically moving charges.
Lepton-pair production also gives access to the proton GPDs via the TCS process.
Another way to  probe the  photon flux is to measure $\eta_c$ production for which
the production rate in $pp$ collisions at AFTER@LHC does depend on the minimal
impact parameter used for the UPC.

The predictions which we obtained for the cross
section for BH -- using specific cuts relevant for the GPD extraction -- 
are on the order of few thousand of pb for the $p$Pb and Pb$p$ cases
and a slightly less than 10 pb for the $pp$ case and we confirm the dominance of BH over TCS.  
This  dominance can partially be overcome by studying of 
the interference -- also sensitive on the GPDs -- between  TCS and BH 
which we evaluated at NLO. With specific cuts on 
the lepton polar angle, the ratio of this interference over the BH amplitude squared
is on order of 10\% with two models of GPDs, \ie\  GK and G-MSTW.
These are quite promising values giving hope for 
the extraction of this interference by means of the analysis of the azimuthal distribution of the produced leptons.
Studying TCS in ultraperipheral collisions at a fixed-target experiment can also give 
us opportunity to study target polarization asymmetries, which are an useful tool 
to extract further information on GPDs \cite{Boer:2015hma}.

We have also derived cross sections for $\eta_c$ production by photon-pair fusion, which
happens, in particular in this energy range, to be sensitive on the method used to compute the photon flux in the $pp$ case. 
We found out that 10$^4$ $\eta_c$  can be produced
per year in UPCs with AFTER@LHC.

Finally, we discussed possible competing hadronic processes via pomeron or odderon exchanges which
could interestingly be separated out by a careful analysis of the transverse-momentum 
dependence of the produced particles.

In conclusion,  AFTER@LHC offers a realistic possibility to study
lepton-pair production in ultraperipheral collisions which opens the path to investigate
features of the partonic structure of hadrons which are complementary to
those studied with lepton beams. The use of hadron beams may, for instance, offer 
the opportunity study to odderon-sensitive reactions.

\section*{Acknowledgements} 
We thank D. d'Enterria, V.P. Gon\c calves, S. Klein, R. Mikkelsen, J. Nystrand for useful discussions. This work is partly supported 
by the COPIN-IN2P3 Agreement, the Polish Grant NCN No DEC-2011/01/D/ST2/02069,  and the CNRS grants
PICS-06149 Torino-IPNO \& PEPS4AFTER2.


\appendix

\section{Kinematics}
\label{appendix-1}
We denote the momenta of incoming nucleons (in the nucleon-nucleon cms) as:
\begin{eqnarray}
p_A &=& \frac{\sqrt{s}}{2}~(1,0,0,\alpha),\nonumber \\
p_B &=& \frac{\sqrt{s}}{2}~(1,0,0,-\alpha),
\end{eqnarray}
where the $A$ is a nucleon from the beam, $B$ is a nucleon from target and $\alpha = \sqrt{1-\frac{4M^2}{s}}$. The Weizs\"acker - Williams photon is emitted from a beam ($\epsilon = -1$) or a target nucleon ($\epsilon = +1$), and its momentum is given by:
\begin{equation}
k = x_\gamma \frac{\sqrt{s}}{2}~(1,0,0,-\epsilon).
\end{equation}
Momentum of outgoing lepton pair (or outgoing virtual photon decaying into heavy lepton pair) reads:
\begin{equation}
q = (q_0,q_T,q_z) \equiv (m_T \cosh y^{\ell\ell}, q_T,m_T \sinh y^{\ell\ell}),
\end{equation}
where $m_T =\sqrt{{q_T}^2 +Q^2}$ and $y^{\ell\ell}$ is the lepton-pair rapidity which can be expressed as
\begin{equation}
y^{\ell\ell} = \frac{1}{2}\epsilon \log \left(
\frac{(Q^2-t)(\alpha+1)}{Q^2(\alpha-1)-t(\alpha-1-2x_\gamma)+sx_\gamma^2(\alpha+1)}
\right), \label{eq:rapidity}
\end{equation}
where: 
\begin{equation}
t \equiv (k-q)^2 = Q^2 -m_T\sqrt{s}x_\gamma e^{\epsilon y^{\ell\ell}} \quad. \label{eq:t}
\end{equation}
Inverting \ce{eq:rapidity}, we can express $x_\gamma$ as a function of $y^{\ell\ell}$:
\begin{eqnarray}
&&x_\gamma(y^{\ell\ell},t,Q,\epsilon) =\\&& \frac{-2t+\sqrt{4t^2-4s(Q^2-t)(\alpha+1)[(\alpha-1)-(\alpha+1)e^{-2\epsilon y^{\ell\ell}}]}}{2s(\alpha+1)}.\nonumber
\label{eq:x}
\end{eqnarray}
or combining \ce{eq:t} with \ce{eq:x} we easily get:
\begin{eqnarray}
t&=& t(y^{\ell\ell},q_T^2,Q,\epsilon),\\
x_\gamma&=& x_\gamma(y^{\ell\ell},q_T^2,Q,\epsilon),\\
J&=& \frac{d t}{d q_T^2}.
\end{eqnarray}

\section{Compton form factors and generalised parton distributions}
\label{appendix-2}
In this appendix, we give the expressions of the LO hard-scattering kernel $T_i$ appearing in
the expression of the Compton form factor ${\cal H}$, $\tilde{\cal H}$ and $\tilde{\cal E}$. 
At Born order, the hard-scattering kernel associated to the quark GPDs are given by 
\begin{eqnarray}
 T^q_H(x,\eta) &=&e_q^2 
     \left(\frac{1}{-\eta-x-i\epsilon} - \frac{1}{-\eta+x-i\epsilon}\right),
 \nonumber \\
T^q_{\tilde{H}}(x,\eta)&=& e_q^2 \left(\frac{1}{-\eta-x-i\epsilon} + \frac{1}{-\eta+x-i\epsilon}\right), 
 \nonumber \\
T^q_E(x,\eta) &=& e_q^2 \left(\frac{1}{-\eta-x-i\epsilon} - \frac{1}{-\eta+x-i\epsilon}\right),
\end{eqnarray}
and those associated to the gluon GPDs are zero. 
The NLO 
hard-scattering kernels can be  found in~\cite{Pire:2011st,Muller:2012yq}. 

\end{document}